**Utility-Based Dose Optimization Approaches for Multiple-Dose Randomized Trial Designs Accounting for Multiple Endpoints**

Gina D'Angelo[a]*, Guannan Chen[b], Di Ran [a]
[a]Oncology Biometrics, Oncology R&D, AstraZeneca, Gaithersburg, Maryland
[b]Department of Statistics, The George Washington University, Washington, DC

*Corresponding author: Gina D'Angelo, AstraZeneca, One MedImmune Way, Gaithersburg, MD 20878;

email: gina.dangelo@astrazeneca.com

**Abstract**

The initiation of dose optimization has driven a paradigm shift in oncology clinical trials to determine the optimal biological dose (OBD). Early-phase trials with randomized doses can facilitate additional investigation of the identified OBD in targeted populations by incorporating safety, efficacy, and biomarker data. To support dose comparison in such settings, we propose to extend the utility score-based approach (U-MET) and introduce the clinical utility index-based approach (CUI-MET) to account for multiple endpoints and doses. The utility-based dose optimization approach for multiple-dose randomized trial designs accounting for multiple endpoints and doses (U-MET-m) extends the U-MET, using a utility score to account for multiple endpoints jointly (e.g., toxicity-efficacy trade-off), while the CUI-MET uses a utility index to do this marginally. U-MET-m and CUI-MET use Bayesian inference within a hypothesis framework to compare utility metrics across doses to identify the OBD. Here we describe simulation studies and present an example to compare the U-MET-m design, CUI-MET, and empirical design. The U-MET-m design and CUI-MET were shown to have satisfactory operating characteristics for selecting the OBD. Based on these findings, we recommend using the U-MET-m and CUI-MET designs as the primary dose comparison approach or as supportive evidence to select the OBD.

# 1 Introduction

Project Optimus is an initiative by the U.S. Food and Drug Administration (FDA) to reform the approaches used to optimize and select doses in oncology drug development. The project has initiated a new framework in oncology to select a dose to take forward to later-phase trials (FDA 2023). Traditional dose-finding approaches have focused on determining the maximum tolerated dose (MTD) by using escalating doses until intolerability is reached, then expanding that dose to evaluate initial efficacy. With the shift from nonspecific chemotherapeutic agents to targeted therapies, however, we can no longer adhere to this dependence on the MTD because it often results in substantial dose discontinuations or reductions of targeted therapies during later phase 3 studies. The FDA is advocating for new approaches that can incorporate a totality of data and evaluate more than a single dose for a longer duration than one dose-limiting toxicity cycle (FDA 2023).

With this shift from the conventional one-dose-fits-all approach to a dose optimization strategy, new approaches are needed to determine the optimal dose, as well as to compare doses and show sufficient evidence from a single-stage design. The FDA suggests the use of dose expansion with randomized doses to explore more doses and enable an unbiased comparison.

Many of the dose optimization methods that have been developed are two-stage where dose escalation and dose expansion are combined. Examples are DROID (Guo and Yuan 2023), MATS (Jiang et al. 2023), MERIT (Yang et al. 2024), U-BOIN (Zhou, Lee, and Yuan 2019), and SDDO (Li et al. 2024). DROID (Guo and Yuan 2023) includes a stage that identifies the therapeutic dose range and the recommended phase 2 dose set, followed by dose expansion to determine the optimal dose. The optimal biologic dose (OBD) is selected based on the Bayesian dose-response index to assess the dose-response relationship. The MATS design (Jiang et al. 2023) evaluates multiple indications from dose escalation and selects doses to include in randomized dose expansion and dose comparison studies, using a Bayesian hierarchical model and information borrowing. MERIT (Yang et al. 2024) is a two-stage approach that identifies the OBD admissible set and determines the sample size. The utility-based Bayesian optimal interval (U-BOIN) (Zhou, Lee, and Yuan 2019) phase 1/2 design is similar to BOIN12, having the aim of determining an OBD by jointly modeling binary toxicity and efficacy endpoints, but is conducted in two stages. SDDO (Li et al. 2024) is a seamless phase 2/3 design that includes randomized doses for dose optimization, an interim analysis, trial decisions, and sample size re-estimation. The U-MET (utility-based dose optimization approach for multiple-dose randomized trial designs) is a single-stage dose optimization approach focused on dose expansion that was developed to compare two doses and include two endpoints (D'Angelo et al. 2024). Although each of the two-stage approaches have merit, are efficient in design, and overlapping objectives, few of these approaches meet the need for a single-stage dose optimization strategy that can determine the OBD at the end of a clinical trial. An advantage to the single-stage dose-optimization approach is the flexibility to adopt any kind of dose-escalation trial and analysis separate from dose expansion. U-MET is the only approach that fits our exact objective and can shed light on which dose is superior to the other doses in terms of efficacy and safety.

The FDA advocates comparing multiple doses and including more data to make an informed dose decision. The totality of data can include a number of endpoints, such as pharmacokinetics and pharmacodynamics, safety, efficacy, quality of life, and nonclinical data (FDA 2023). Approaches that incorporate multiple endpoints are needed to address the totality of data.

Here we describe our extension of the U-MET approach (D'Angelo et al. 2024) to allow comparison of the standardized mean utilities across multiple doses and multiple endpoints. We refer to this extended approach as U-MET-m (U-MET accounting for multiple endpoints and doses). U-MET and U-MET-m combine the endpoints jointly and assign weights to different combinations of the endpoints. We also introduce the CUI-MET (clinical utility index dose optimization approach for multiple-dose randomized trial designs) which is a simple and natural approach to combine data, where it combines endpoints marginally with a utility index. Although there is an advantage to assigning weights jointly this becomes more complicated to assign as the number of endpoints increases and will become too difficult beyond three endpoints, pointing to the advantage of the CUI-MET in this case. Both the U-MET-m and CUI-MET approaches require thresholds for the endpoints to create binary variables that indicate a patient's response.

In this manuscript, we extended the U-MET approach specifically to compare more than two doses and incorporate three endpoints. The higher the standardized mean utility, the more desirable the dose. Assuming that the lower dose is safer, the aim was to determine whether higher doses "substantially" increase efficacy without impacting safety, thereby identifying a tolerable dose that maximizes efficacy. CUI-MET extends the Bayesian inference from U-MET to enable the comparison of doses with utility indices. We focused on a decision-making approach to assess the strength of evidence and select the "best" dose from randomized dosed arms in settings with multiple doses and/or endpoints.

In the methods section, we will describe the U-MET-m, CUI-MET, and empirical design. We describe simulation studies and a hypothetical oncology trial example to demonstrate the operating characteristics of the approaches. We close with a discussion.

## 2 Methods
### 2.1 U-MET design

In the U-MET design described previously by D'Angelo et al. (D'Angelo et al. 2024), two doses from dose expansion studies were compared with two endpoints. The primary question addressed in the U-MET design is whether a higher dose is superior to a lower dose in a dose expansion trial in which doses are randomized. To address this question, the BOIN12 utility methodology was extended to a hypothesis-testing framework with a Bayesian approach. BOIN12 (Lin et al. 2020) is a dose-finding approach that is used to select the optimal dose by accounting for both safety and efficacy data in a (phase 1/2) dose escalation trial. To this end, BOIN12 uses utilities to measure the efficacy-toxicity tradeoff (i.e., desirability) of a dose and makes the dose escalation/de-escalation decision based on a ranking of the doses' desirability. For our purposes, we defined a binary toxicity endpoint and a binary efficacy endpoint that produce four possible efficacy-toxicity outcomes: (1) efficacy, no toxicity; (2) no efficacy, no toxicity; (3) efficacy, toxicity; and (4) no efficacy, toxicity. The utility scores $(u_1, u_2, u_3, u_4)$ can be obtained from the clinical team to reflect the desirability of the four possible outcomes (see Table 1).

Table 1: Utility scores

| Toxicity/Endpoint 2 | Efficacy/Endpoint 1 | |
|---|---|---|
|  | Yes | No |
| No | $u_1$ | $u_2$ |
| Yes | $u_3$ | $u_4$ |

It is recommended that the utility score of the most desirable outcome is set as $u_1 = 100$ and the utility score of the least desirable outcome is set as $u_4 = 0$ as a reference, while the utility scores associated with the other two outcomes (i.e., $u_2$ and $u_3$) are specified as $(u_2, u_3) \in [0,100]$ (Lin et al. 2020).

Given a set of utility scores $(u_1, u_2, u_3, u_4)$, the mean utility of dose $d$ is calculated as a weighted average of $(u_1, u_2, u_3, u_4)$ by their corresponding probabilities, as follows:

$$u(d) = \sum_{k=1}^{4} p_k(d) u_k,$$

where $p_k(d)$ is the probability of efficacy-toxicity outcomes $k$, $k = 1,\ldots,4$, for dose $d$. A special case that simplifies the calculation of $u(d)$ occurs when $u_2 + u_3 = 100$. In this case, $u(d) = u_2(1 - p_T(d)) + u_3 p_E(d)$ (Lin et al. 2020), where $p_T(d)$ is the marginal probability of toxicity and $p_E(d)$ is the marginal probability of efficacy for dose $d$. This implies that to calculate $u(d)$, we only need summary data statistics (i.e., marginal probabilities) for efficacy and toxicity, rather than subject-level data or probabilities for each efficacy-toxicity outcome.

Following BOIN12 and U-MET, we used a quasi-binomial likelihood method to make inference of $u^*(d)$ and define the standardized mean utility $u^*(d) = u(d)/100$. Therefore, $u^*(d)$ can be regarded as a probability and modeled using a binomial distribution with "quasi-binomial" data $(x(d), n(d))$, as follows (Lin et al. 2020):

$$x(d)|u^*(d) \sim Binomial(n(d), u^*(d)),$$

where $n(d)$ denotes the number of subjects treated at dose $d$, $y_k(d)$ denotes the number of subjects with efficacy-toxicity outcome $k$, and $x(d) = \frac{u_1 y_1(d) + u_2 y_2(d) + u_3 y_3(d) + u_4 y_4(d)}{100}$ represents the number of "quasi" events among $n(d)$ subjects. When $u_2 + u_3 = 100$, $x(d) = \frac{u_2(n(d) - n_T(d)) + u_3 n_E(d)}{100}$, where $n_T(d)$ is the number of patients with a toxicity and $n_E(d)$ is the number of patients with a response (efficacy event).

BOIN12 and U-MET utilize a Bayesian approach, where the prior is $u^*(d) \sim Beta(\alpha, \beta)$, resulting in the posterior distribution of $u^*(d)$ given by

$$u^*(d)|D(d) \sim Beta(\alpha + x(d), \beta + n(d) - x(d)),$$

where $D(d)$ is the observed data. The hyperparameters $\alpha$ and $\beta$ are often set as 1 to obtain a non-informative prior.

## 2.2 U-MET-m design

Now that we have introduced the framework around $u^*(d)$, we resume the discussion of extending the U-MET methodology to a hypothesis-testing framework with a Bayesian approach for multiple doses, and in a later section account for multiple endpoints. The higher dose is referred to as "dose $h$" and the lower dose as "dose $l$" where these terms are used interchangeably and $1 \leq l < h$. We also refer to the difference as Δ and diff, using these terms interchangeably. Define the standardized mean utility difference between the higher and lower doses as $u_\Delta^* = u^*(h) - u^*(l)$, where $u^*(l)$ and $u^*(h)$ denote the standardized mean utility for the lower dose and the higher dose, respectively. We consider the following hypotheses:

$$H_0: u_\Delta^* \leq \delta \text{ vs. } H_1: u_\Delta^* > \delta,$$

where $\delta$ is a predefined difference, which is often set to 0. However, the $\delta$ can be any clinically meaningful difference.

### 2.2.1 Admissible criteria

It is advisable to eliminate doses that are considered inadmissible and to exclude them from the comparison. Dose level $d$ is defined as admissible if the data indicate that $d$ is reasonably safe and efficacious. The safety and efficacy rules are as follows:

$$\text{Toxic (safety rule): } Pr(p_T(d) > \phi_T | D(d)) > c_T,$$

$$\text{Futile (Efficacy rule): } Pr(p_E(d) < \phi_E | D(d)) > c_E,$$

where $\phi_T$ is the toxicity upper limit, $\phi_E$ is the efficacy lower limit, $c_T$ and $c_E$ are probability cut-offs for toxicity and efficacy, and $D(d) = (n(d), y_1(d), y_2(d), y_3(d), y_4(d))$ is the observed data at dose level $d$. If more endpoints are considered, they can be added to the criteria, although this is not required.

### 2.2.2 U-MET-m design using Bayesian decision rule

The U-MET-m design evaluates the hypotheses $H_0$ and $H_1$ and makes a decision on dose selection. This dose decision is based on $p(u_\Delta^* > 0)$, i.e., the probability that the standardized mean utility for dose $h$ is strictly greater than the standardized mean utility for dose $l$, $p(u^*(h) > u^*(l))$. A more general decision rule might instead evaluate $p(u_\Delta^* > \delta)$ for a minimally important difference $\delta$, where a minimally important difference between the standardized mean utilities would be evaluated.

Specifically, as described by Sverdlov et al. (Sverdlov, Ryeznik, and Wu 2015), the cumulative distribution function of the difference between $u^*(h)$ and $u^*(l)$ is

$$F_{u^*_\Delta}(t) = Pr(u^*_\Delta \leq t) = Pr(u^*(h) - u^*(l) \leq t)$$

$$= \begin{cases} \int_{-t}^{1} F_{u^*(h)}(t+v) f_{u^*(l)}(v) dv, -1 \leq t \leq 0 \\ \int_{0}^{1-t} F_{u^*(h)}(t+v) f_{u^*(l)}(v) dv + \int_{1-t}^{1} f_{u^*(l)}(v) dv, 0 \leq t \leq 1 \end{cases},$$

where $f_{u^*(d)}(v)$ is the posterior density function (i.e., a beta pdf) and $F_{u^*(d)}$ is the posterior cumulative distribution function for $u^*(d)$.

Let $C_1$ and $C_2$ denote prespecified probability cutoffs, with $C_2 < C_1$. The U-MET-m design makes the dose selection decision as follows:

- Select the high dose if $p(u^*_\Delta > 0) = 1 - F(0) > C_1$.
- Select the low dose if $p(u^*_\Delta > 0) < C_2$.
- Otherwise, make a "consider high dose" decision if $C_2 \leq p(u^*_\Delta > 0) \leq C_1$.

We will select $C_1$ to be (1- $\alpha_1$) and $C_2$ to be (1- $\alpha_2$). Refer to D'Angelo et al. (D'Angelo et al. 2024) for discussion of $\alpha_1$ and $\alpha_2$.

When there are more than two doses, we offer two strategies, a sequential testing strategy without "consider zones" and a pairwise testing strategy with consider zones, to arrive at a final dosing decision. The sequential testing strategy compares all doses to the highest admissible dose in a stepwise manner while removing the consider zone to simplify the process. The pairwise testing strategy, on the other hand, evaluates all pairwise comparisons between doses, retaining the consider zone to allow users more flexibility in the final dose selection.

The sequential dose selection decision without consider zones is as follows:

- Select the high dose if $p(u_\Delta^* > 0) = 1 - F(0) > C_1$.

- Otherwise, select the low dose if $p(u_\Delta^* > 0) <= C_1$.

This strategy starts with the dose that has the highest desirability and compares it to the lowest dose. If the lowest dose is selected, the testing stops. This assumes that any other doses between the highest and lowest doses do not provide a meaningful improvement. If there is a difference between the lowest and highest dose, testing proceeds by comparing the next-lowest dose to the highest dose. This process continues until it is determined that the highest dose is substantially different from all lower doses, or until any lower dose is found to be not substantially different from the highest dose.

The algorithm for the sequential testing without the consider zone is as follows:

1. Begin with the randomized dose arms (dose 1, dose 2, …, dose D), where $d$ = 1, …, D denotes dose level.

2. Identify the most desirable dose: determine the dose $d^*$ with the highest desirability.

3. Compare dose $d^*$ to all the lower doses ($d$ = 1, …, $d^* - 1$), starting with the lowest dose and proceeding to the next-higher dose until a difference is no longer found or all doses differ from dose $d^*$.

4. Summary: This algorithm continues until
    a. a lower dose is found to not differ from dose $d^*$, and that dose is selected, or
    b. dose $d^*$ differs from all doses below it, and dose $d^*$ is selected.

### 2.2.3 U-MET-m for three endpoints

When a third endpoint is incorporated, Table 1 would need to be further broken down by the third endpoint. For example, in addition to efficacy and safety, we may incorporate a (PD) biomarker as a third endpoint. The BOIN12 approach can incorporate three endpoints to make a dose escalation decision (Lin et al. 2020), which we adapt here for our extension of U-MET-m. Table 2 shows the utility score assignments for three endpoints.

| Table 2: Utility score for three endpoints | | | | |
|---|---|---|---|---|
| | Third endpoint (Response, aka PD Biomarker) | | | |
| | Yes | | No | |
| | Efficacy | | Efficacy | |
| | Yes | No | Yes | No |
| Toxicity | | | | |
| No | $u_{11}$ | $u_{12}$ | $u_{01}$ | $u_{02}$ |
| Yes | $u_{13}$ | $u_{14}$ | $u_{03}$ | $u_{04}$ |

Let $p_{jk}(d)$ denote the probabilities of the biomarker level $j$ and efficacy-toxicity outcome $k$ associated with $u_{jk}$ at dose level $d$, where $j = 0,1$, $k = 1,..,4$, and $d = 1,..,D$. The desirability score at dose $d$ is $u(d) = \sum_{j=0}^{1} \sum_{k=1}^{4} p_{jk}(d) u_{jk}$. We let $y_{jk}(d)$ be the number of patients experiencing the outcomes associated with $u_{jk}$ at dose level $d$. The observed desirability at dose $d$ is

$$\hat{u}(d) = \frac{\sum_{j=0}^{1} \sum_{k=1}^{4} y_{jk}(d) u_{jk}}{n(d)},$$

where $n(d)$ is the number of subjects at dose $d$. The multivariate outcome can again be converted into a univariate desirability outcome. The dose comparison follows the same algorithm described in section 2.2.2.

## 2.3 CUI-MET design

We propose the use of a clinical utility index (CUI) as a composite endpoint to combine data from multiple endpoints to choose an optimal dose (Ouellet et al. 2009; Ouellet 2010; Winzenborg, Soliman, and Shebley 2021). Composite endpoints combine two or more endpoints into a single variable. This can be done in various ways, such as using BOIN12, U-MET, and CUI. CUI integrates the benefit-risk tradeoff by providing weights to each endpoint and combining the endpoints marginally (Ouellet et al. 2009; Ouellet 2010), whereas utility score–based approaches (e.g., BOIN12, U-MET, U-MET-m) combine the endpoints jointly. The CUI can be calculated based on a weighted average of the endpoints, for example, an arithmetic weighted average of endpoints (Song et al. 2013; Winzenborg, Soliman, and Shebley 2021) or a geometric mean of a 0–1 desirability score calculated for each endpoint (Coffey, Gennings, and Moser 2007). The directional effect of each endpoint should be the same for the CUI to be sensible (e.g., when including response and toxicity, the higher values should indicate better patient outcomes, such as ORR = 1 and no toxicity = 1). In this work, the CUI is estimated empirically and compared across doses to make a dosing decision.

For multiple endpoints, $CUI(d) = \sum_{m=1}^{M} p_m(d) w_m$, where $p_m$ is the marginal probability of observing the $mth$ endpoint at dose $d$ and $w_m$ is the weight for the $mth$ endpoint satisfying $\sum_{m=1}^{M} w_m = 1$. Also note that $0 \leq CUI(d) \leq 1$. A specific case with two endpoints of toxicity and safety results in $CUI(d) = (1 - p_T(d))w_T + p_E w_E(d)$, where $T$ and $E$ indicate toxicity and efficacy, respectively.

In a similar fashion to U-MET-m, $CUI(d)$ can be regarded as a probability and can be modeled using a binomial distribution with "quasi-binomial" data $(x_{cui}(d), n(d))$ as follows:

$$x_{cui}(d)|CUI(d) \sim Binomial(n(d), CUI(d)),$$

where $x_{cui}(d) = n(d) \sum_{m=1}^{M} p_m(d) w_m$ represents the number of "quasi" events among $n(d)$ subjects. When there are only two endpoints, with $u_2 + u_3 = 100$ for U-MET-m, $\frac{u_2}{100} = w_1$, and $\frac{u_3}{100} = w_2$, CUI and U-MET-m will be equivalent.

As with U-MET-m, we take a Bayesian approach for CUI-MET. Assuming the prior $CUI(d) \sim Beta(\alpha, \beta)$, the posterior distribution of $CUI(d)$ is given by

$$CUI(d)|D(d) \sim Beta(\alpha + x_{cui}(d), \beta + n(d) - x_{cui}(d)),$$

where $D(d)$ is the observed data. The hyperparameters $\alpha$ and $\beta$ are often set as 1 to obtain a non-informative prior. Here we have the clinical utility difference as $CUI_\Delta = CUI(h) - CUI(l)$, and the same dose comparison schema will follow for $CUI_\Delta$ as for $u_\Delta^*$. We can apply the same approach used for CUI-MET to U-MET-m to compare doses, using the exact Beta distribution from section 2.2.2.

## 2.4 Relate U-MET-m to CUI-MET

To explore how U-MET-m can be related to CUI-MET, we first consider a scenario in which there are two endpoints followed by three endpoints. Suppose we have two endpoints as described in section 2.1. Let $p_k(d)$ denote the probabilities of the efficacy-toxicity outcome $k$ associated with $u_k$ at dose level $d$, $d = 1,..,D$, and $k = 1,..,4$. The desirability score at dose $d$ is $u(d) = \sum_{k=1}^{4} p_k(d) u_k$.

We make the following assumptions for the utility score and probabilities:

$$u_1 = 100, u_2 + u_3 = 100, u_4 = 0 \text{ and } 1 - p_T = (p_1 + p_2), p_E = (p_1 + p_3),$$

where $T$ and $E$ indicate toxicity and efficacy, respectively. Under these assumptions, the desirability score simplifies to:

$$u(d) = p_1 \times 100 + p_2 u_2 + p_3 u_3 + p_4 \times 0$$
$$= p_1 (u_2 + u_3) + p_2 u_2 + p_3 u_3$$
$$= (1 - p_T) u_2 + p_E u_3.$$

The CUI-MET is $CUI(d) = (1 - p_T) w_T + p_E w_E$.

The U-MET-m utility ($u(d)$) and CUI ($CUI(d)$) are equivalent when $u_2/100 = w_T$ and $u_3/100 = w_E$, and the above assumptions are made for the utility score where $u_2 + u_3 = 100$.

Now suppose we have a third endpoint (see Table 2 for the utility scores). We are unsure of the direction of the third endpoint, but for demonstrative purposes, the third endpoint is a biomarker, and here the positive biomarker is the "preferred" and "efficacious" directions. The CUI is $CUI(d) = (1 - p_T)w_T + p_E w_E + p_B w_B$, where $T$, $E$, and $B$ indicate toxicity, efficacy, and biomarker, respectively. For our purposes here, the third endpoint is referred to as a biomarker, although this can be any type of endpoint, such as tolerability, quality of life, etc.

Let $p_{jk}(d)$ denote the probabilities of biomarker level $j$ and efficacy-toxicity outcome $k$ associated with $u_{jk}$ at dose level $d, d = 1, .., D, j = 0,1$, and $k = 1, ..,4$. The desirability score at dose $d$ is

$$u(d) = \sum_{j=0}^{1} \sum_{k=1}^{4} p_{jk}(d) \, u_{jk}$$

$$= p_{11}u_{11} + p_{12}u_{12} + p_{13}u_{13} + p_{14}u_{14} + p_{01}u_{01} + p_{02}u_{02} + p_{03}u_{03} + p_{04}u_{04}.$$

We know that:

$$p_E = p_{11} + p_{13} + p_{01} + p_{03},$$

$$1 - p_T = p_{11} + p_{12} + p_{01} + p_{02},$$

$$p_B = p_{11} + p_{12} + p_{13} + p_{14}.$$

Below is general guidance on how to select utility scores for the biomarker-positive and -negative side that can lend to clinical interpretation. We may assume a constant $(a_1, .., a_4)$, where $u_{1i} \geq a_i \geq 0, i = 1, ..4$ is added to the utility score from the biomarker-negative side to provide the utility scores for the biomarker-positive side. This can down-weight the biomarker-negative side, which would represent that the biomarker-positive patients indicate biological activity of the drug or that the patient may respond to treatment if they have that biomarker. If we assume:

$$u_{11} = u_{01} + a_1, u_{11} \geq a_1 \geq 0,$$

$$u_{12} = u_{02} + a_2, u_{12} \geq a_2 \geq 0,$$

$$u_{13} = u_{03} + a_3, u_{13} \geq a_3 \geq 0,$$

$$u_{14} = u_{04} + a_4, u_{14} \geq a_4 \geq 0,$$

Then $u(d) = (u_{11}p_{11} + u_{12}p_{12} + u_{13}p_{13} + u_{14}p_{14}) + (u_{01}p_{01} + u_{02}p_{02} + u_{03}p_{03} + u_{04}p_{04})$

$= (u_{01}p_{11} + u_{02}p_{12} + u_{03}p_{13} + u_{04}p_{14}) + (u_{01}p_{01} + u_{02}p_{02} + u_{03}p_{03} + u_{04}p_{04}) +$

$(a_1 p_{11} + a_2 p_{12} + a_3 p_{13} + a_4 p_{14})$.

To demonstrate that a relationship can exist between U-MET-m and CUI-MET, we can consider a few scenarios. The one condition that must be met is $0 \leq u_{jk} \leq 100$.

For the first scenario, suppose $u_{j1} = u_{j2} + u_{j3}$, $a_1 = a_4 = a$, $a_2 = a_3 = a/2$. That is, we have: $u_{11} = u_{12} + u_{13}$, $u_{01} = u_{02} + u_{03}$, $u_{12} = u_{02} + \frac{a}{2}$, $u_{13} = u_{03} + a/2$, $u_{14} = a$, and $u_{04} = u_{14} - a$. This calculates $u(d)$ to be

$$u(d) = (1 - p_T)u_{02} + p_E u_{03} + p_B a - (p_{12} + p_{13}) * \frac{a}{2} + p_{04} u_{04}.$$

This shows that $u(d)$ is similar to $CUI(d) = (1 - p_T)w_T + p_E w_E + p_B w_B$ under the conditions stated above, and let $w_T = \frac{u_{02}}{100}$, $w_E = \frac{u_{03}}{100}$, and $a = w_B$, where the weights add up to 1. As $a$ increases, the more the biomarker-negative side is down-weighted, and the biomarker weight would increase as a result of CUI. A special case may be letting $u_{04} = 0$ or $u_{14} = u_{04} = 0$.

A second scenario is to let $a_1 = a_2 = a_3 = a_4 = a$, the same constant for all toxicity-efficacy outcomes. Then $u(d) = (1 - p_T)u_{02} + p_E u_{03} + p_B a$, which is equivalent to $CUI(d)$ when $w_T = \frac{u_{02}}{100}$, $w_E = \frac{u_{03}}{100}$, and $w_B = a$. As $a$ increases, the biomarker-negative side is even more down-weighted than in the first scenario. A special case may be letting $u_{04} = 0$ or $u_{14} = u_{04} = 0$.

In these scenarios, $a$ would need to be selected such that none of the weights are negative. When the weights are kept the same for the biomarker-positive and -negative side for U-MET-m, $a_i = 0 = a$, the biomarker will have no impact on the utility values. However, it is not a condition to select $a$ based on $u(d)$ and $a_i$; rather, this shows how to relate U-MET-m to CUI-MET. Although we have shown that there can be a relationship between U-MET-m and CUI-MET, the former accounts for the weights jointly, and as $a$ increases, the U-MET-m mean utility differs from CUI(d).

## 2.5 Empirical design

To enable the assessment of the operating characteristics of the U-MET-m and CUI-MET designs, we selected an empirical design in which dose selection is based on the observed toxicity ratio (TR) of the high dose compared with that of the low dose and the observed efficacy rate differences (EDs) of the high dose is compared with those of the low dose. Table 3 demonstrates this empirical design, which requires the specification of two thresholds for toxicity (TR1, TR2) and two thresholds for efficacy (ED1, ED2). By comparing ED and TR to these thresholds, a decision is made to select the low dose, to consider the high dose, or to select the high dose. For example, if the ED is greater than the threshold ED2 (ED > ED2) and the TR is less than TR1 (TR < TR1), then we select the high dose.

| Table 3. Empirical design based on marginal efficacy and toxicity probabilities | | | |
|---|---|---|---|
| Toxicity ratio | Efficacy difference | | |
| | >$ED_2$ | [$ED_1$,$ED_2$] | < $ED_1$ |
| <$TR_1$ | High | High | Consider* |
| [$TR_1$, $TR_2$] | High | Consider | Low |
| >$TR_2$ | Consider | Low | Low |
| Toxicity ratio (TR) is the ratio of the toxicity of the higher dose to that of the lower dose ($Tox_H$/$Tox_L$). Toxicity thresholds are $TR_1$ and $TR_2$. Efficacy difference (ED) is the difference between the efficacy of the higher dose and that of the lower dose ($Eff_H$-$Eff_L$). Efficacy thresholds are $ED_1$ and $ED_2$. *An alternative dose selection for the scenario of ED<0 can be Low dose, and this will be indicated a priori. | | | |

The empirical design uses different criteria from U-MET-m and CUI-MET to determine the efficacy-toxicity trade-off for dose selection. Whereas U-MET-m and CUI-MET incorporate utility, the empirical design does not have an explicit trade-off metric but rather relies on the empirical decision rule. Our objective in including this empirical design is to use it as a reference to better understand the operating characteristics of U-MET-m and CUI-MET. Therefore, any direct comparison between these designs should be interpreted with caution. In addition, because the decision table assumes that the efficacy of the lower dose will always be less than that of the higher dose, it might be advisable to perform an additional check when ED < 0 so that the lower dose would be selected. Table 4 shows how the empirical design accounts for a third endpoint, where the third endpoint may be a "tie-breaker" for the consider zone.

| Table 4. Empirical design based on marginal efficacy, toxicity, and biomarker probabilities | | | | | | |
|---|---|---|---|---|---|---|
| | Biomarker difference>$BD_1$ | | | Biomarker difference<=$BD_1$ | | |
| Toxicity ratio | Efficacy difference | | | Efficacy difference | | |
| | >$ED_2$ | [$ED_1$,$ED_2$] | < $ED_1$ | >$ED_2$ | [$ED_1$, $ED_2$] | < $ED_1$ |
| <$TR_1$ | High | High | High | High | High | Consider* |
| [$TR_1$, $TR_2$] | High | Consider | Low | High | Consider | Low |
| >$TR_2$ | Consider | Low | Low | Low | Low | Low |
| Toxicity ratio (TR) is the ratio of the toxicity of the higher dose to that of the lower dose. Toxicity thresholds are $TR_1$ and $TR_2$. Efficacy difference (ED) is the difference between the efficacy of the higher dose and that of the lower dose. Efficacy thresholds are $ED_1$ and $ED_2$. Biomarker difference (BD) is the difference between the biomarker levels in the higher dose and those in the lower dose. Biomarker threshold is $BD_1$. *An alternative dose selection for the scenario of ED<0 can be low dose, and this will be indicated a priori. | | | | | | |

# 3 Simulation
## 3.1 Simulation set-up

We conducted simulation studies to evaluate the operating characteristics of the U-MET-m design compared with the CUI-MET and the empirical design. This comparison is crucial as, at present, dose selection often relies heavily on clinical judgment without a robust statistical approach. We avoided a comparison of U-MET-m and CUI-MET with other two-stage designs, such as DROID, because our approach is a single-stage approach, whereas DROID contains two stages (namely, the dose-finding stage and the randomization stage). To assess the characteristics, the following values were fixed across all scenarios: the lower threshold of efficacy was $\phi_E = 0.22$, the upper threshold of toxicity was $\phi_T = 0.35$, and the admissible criteria was $c_T = 0.95$, and $c_E = 0.90$. Each simulation study had 1,000 replications.

The first simulation study compared the methods to the truth described below. We calculated the dose selected for each simulation study. To assess the characteristics, the following values were fixed across all scenarios: n = 40/arm, low-dose efficacy rate $p_{E1} = 0.4$, low-dose toxicity rate $p_{T1} = 0.13$, and utility scores ($u_1 = 100, u_2 = 35, u_3 = 65, u_4 = 0$) with the various high-dose efficacy and toxicity rates specified in Table 5. For the high dose, $p_{E2}, p_{T2}$, and the "true" outcome were derived using a binomial test with a large enough difference ($\alpha = 0.2$)/borderline difference ($\alpha = 0.34$)/no difference between the high and low doses for efficacy and toxicity pre-defined in Table 5 (scenarios 1 through 9) to then make a decision regarding high or low dose in the sequential testing procedure. Table 3 shows the empirical design outcomes when we considered ED cutoffs ED1 = 0.15 and ED2 = 0.35 and TR thresholds TR 1 = 1.5 and TR2 = 2 for the empirical design decision (Table S1). For notational simplicity in this and the next sections regarding dose notation, L = low dose and H = high dose.

The second set of simulation studies had the truth based on scenarios set for the empirical design, U-MET-m, and CUI-MET approaches with two endpoints (Tables 6 and S2). When $u_2 + u_3 = 100$ with two endpoints, U-MET-m and CUI were the same, and we do not display the CUI-MET results in these cases. When there are three endpoints, all three approaches are reported (Tables 7, S3, and S4). The truth differed across approaches, which demonstrates the differences among the methods. This set of simulations had fixed lower-dose efficacy and toxicity rates and selected representative higher-dose efficacy and toxicity rates from each cell of Table 3 (Table S1) so that we could compare the dosing decisions of the empirical design to those of the U-MET-m. The truth was calculated based on the fixed higher and lower rates for the empirical design and the U-MET-m (and CUI-MET for the above-mentioned scenarios). The simulated results for the empirical design and U-MET-m were compared to their respective truth sets for the empirical design and the U-MET-m, with CUI-MET included for scenarios involving three endpoints.

To assess the characteristics, the following values were fixed across all scenarios: the sample size was *n* = 30 per dosed arm, and utility scores were ($u_1 = 100, u_2 = 35, u_3 = 65, u_4 = 0$). The sample size was reduced here from 40 to 30 to demonstrate the operating characteristics under a smaller sample size, which mimics what might be used in clinical trials. Tables 3 and S1 show the empirical design outcomes when we considered ED cutoffs ED1 = 0.15 and ED2 = 0.35 for the empirical design decision. For the sequential testing procedure, we evaluated two scenarios: (1) when we allowed the consider zone to be low dose for a dosing decision for the empirical table and $\alpha_1 = 0.2$, and (2) when we allowed the consider zone to be high dose for a dosing decision for the empirical table and $\alpha_1 = 0.34$.

Tables 6 and S2 presents the high-dose efficacy and toxicity rates with the low-dose efficacy and toxicity rate of ($p_{E1} = 0.23, p_{T1} = 0.13$) for Table 6, and low-dose efficacy and toxicity rate of ($p_{E1} = 0.40, p_{T1} = 0.13$) for Table S2. Tables 7, S3, and S4 show a third endpoint representing a biomarker with a low rate of 0.2. Table 7 shows high-dose efficacy and toxicity rates with the low-dose efficacy and toxicity rate of ($p_{E1} = 0.23, p_{T1} = 0.13$) when the consider zone was low dose and when it was high dose. Tables S3 and S4 show high-dose efficacy and toxicity rates with the low-dose efficacy and toxicity rate of ($p_{E1} = 0.40, p_{T1} = 0.13,$) when the consider zone was low dose (Table S3) and when it was high dose (Table S4). Table 4 presents the empirical design outcomes when we considered ED cutoffs of ED1 = 0.15 and ED2 = 0.35, TR thresholds of TR1 = 1.5 and TR2 = 2, and a biomarker difference (BD) threshold of BD1 = 0.1 for the empirical design decision. For the results shown in Tables 7, S3, and S4, we used the first scenario described in section 2.4, which relates U-MET-m to CUI-MET where $a_1 = a_4 = 10, a_2 = a_3 = 5$ and utility scores are ($u_{11} = u_{12} + u_{13}, u_{12} = 35, u_{13} = 65, u_{14} = 0, u_{01} = u_{02} + u_{03}, u_{02} = 30, u_{03} = 60, u_{04} = 0$). The small weight of $a_1 = 10$ gives a small weight for the biomarker of 0.10 in CUI-MET. Additional values of $a_i$ were evaluated to assess what happens as $a_i$ increases. We down-weighted the biomarker-negative utility scores because we placed a greater value on the biomarker-positive score. The Supplemental Material shows utility scores that have a different set-up for the three endpoints. Additional details of data generation for the outcomes is provided in the Supplemental Material. For our purposes, we generated the data to have no correlations, and in previous evaluations we found correlation to not have an impact; however, this can be explored in future work.

## 3.2 Simulation results

Results for the true dose comparison are reported in Table 5 and Figures S1 and S2. The U-MET-m produced a higher percentage than the empirical decision in selecting the correct dose when $\alpha_1 = 0.2$ and the consider zone was the low dose (Table 5 and Figure S1). In about half of the scenarios when the highest dose had a much higher efficacy rate, the U-MET-m had at least a 70% chance of selecting the correct dose. When the highest and next-highest doses had closer efficacies, U-MET-m had at least a 50% chance of selecting the correct dose (50–62%). We also evaluated when ED < 0 would be low dose, and the findings were the same [results not presented]. In addition, we evaluated when consider would be high dose and $\alpha_1 = 0.34$ (Figure S2). The results were mostly the same but with an increased percentage of selecting the correct dose. An exception was scenario 2, in which for the truth the consider zone is high dose to make this a fair comparison; in this case, dose 3 would be the truth.

In the second set of simulations, results are presented for two endpoints and for three endpoints. The results for two endpoints will be discussed first. When $\alpha_1 = 0.20$ and the consider zone was the low dose, the true dose determined by U-MET-m tended to be higher than the one produced by the empirical design. Across most scenarios, the U-MET-m was more likely than the empirical design to select the true dose. For the low-dose efficacy and toxicity rate of $(p_{E1} = 0.23, p_{T1} = 0.13)$ (Table 6, Figure 1), U-MET-m performed better than the empirical approach for scenarios 2, 4, 5, 6, and 9 and similarly for scenarios 1 and 7. When $\alpha_1 = 0.34$ and the consider zone was the high dose (Table 6, Figure 2), most of the time the true dose tended to be the highest dose for both approaches; exceptions were scenarios 8 and 9, in which the approaches differed in the dose selection. The U-MET-m was more likely than the empirical design to select the true dose across most scenarios and was about the same in a few scenarios (scenarios 5 and 6).

| Table 5. Simulation results of percentage correct for dose selection compared to the truth[a] | | | | | |
|---|---|---|---|---|---|
| Scenario | High Dose Efficacy rate | High Dose Toxicity rate | True outcome | Emp/ U-MET-m, % correct | Emp/ U-MET-m selected |
| 1 | (0.45,0.55,0.75) | (0.15,0.2,0.28) | Dose 4 | 9/50 | 1/4 |
| 2 | (0.45,0.55) | (0.15,0.2) | Dose 2 | 16/29 | 1/1 |
| 3 | (0.45,0.75) | (0.15,0.28) | Dose 3 | 17/73 | 1/3 |
| 4 | (0.55,0.45) | (0.15,0.2) | Dose 2 | 9/53 | 1/2 |
| 5 | (0.55,0.75) | (0.2,0.28) | Dose 3 | 17/54 | 1/3 |
| 6 | (0.45,0.58,0.8) | (0.17,0.23,0.3) | Dose 4 | 13/62 | 1/4 |
| 7 | (0.58,0.8) | (0.17,0.23) | Dose 3 | 31/70 | 1/3 |
| 8 | (0.45,0.8) | (0.17,0.3) | Dose 3 | 20/84 | 1/3 |
| 9 | (0.58,0.8) | (0.23,0.23) | Dose 3 | 40/78 | 1/3 |
| [a] Fixed parameters: $n = 40$; low dose efficacy rate, 0.4; low dose toxicity rate, 0.13; $(u_2, u_3)$ = (35, 65); $(ED_1, ED_2)$ = (0.15, 0.35), consider is low and $\alpha_1 = 0.2$. Truth: Sequential testing procedure: For each step High dose [Eff diff & Tox (no diff, bord diff)]; Else low dose. Notation: Emp, empirical design; eff, efficacy; tox, toxicity; diff, difference; bord, borderline; anything, (no diff, bord, diff). | | | | | |

| Table 6: Simulation results of percentage correct for dose selection compared to empirical design and U-MET-m truth[a] | | | | | | | | |
|---|---|---|---|---|---|---|---|---|
| | | | $\alpha_1$=0.20, consider is low dose | | | $\alpha_1$=0.34, consider is high dose | | |
| | | | Simulation results: compare to True dose for each approach and report on % correct (list one if the same) | | | Simulation results: compare to True dose for each approach and report on % correct (list one if the same) | | |
| Scenario | High Efficacy | High Toxicity | True dose for Emp/ U-MET-m (list one if the same) | Emp | U-MET-m | True dose for Emp/ U-MET-m (list one if the same) | Emp | U-MET-m |
| 1 | (0.47,0.70) | (0.20,0.28) | 1/3 | 54/25 | 3/55 | 3 | 70 | 77 |
| 2 | (0.27,0.70) | (0.15,0.28) | 1/3 | 55/28 | 5/89 | 3 | 84 | 97 |
| 3 | (0.27,0.47) | (0.15,0.20) | 1/2 | 60/19 | 41/20 | 3 | 55 | 66 |
| 4 | (0.47,0.70) | (0.15,0.20) | 3 | 33 | 62 | 3 | 74 | 82 |
| 5 | (0.47,0.70) | (0.20, 0.20) | 3 | 42 | 69 | 3 | 86 | 87 |
| 6 | (0.47,0.70) | (0.28, 0.28) | 1/3 | 55/32 | 3/66 | 3 | 86 | 85 |
| 7 | (0.27,0.47, 0.70) | (0.15, 0.20, 0.28) | 1/4 | 52/16 | 3/53 | 4 | 66 | 75 |
| 8 | (0.47,0.27) | (0.15, 0.20) | 1/2 | 88/12 | 32/68 | 1/2 | 48/18 | 14/85 |
| 9 | (0.47,0.47) | (0.15, 0.20) | 1/2 | 60/36 | 21/72 | 3/2 | 45/25 | 21/73 |
| [a] Fixed parameters: $n = 30$; low dose efficacy rate, 0.23; low dose toxicity rate, 0.13; $(u_2, u_3)$ = (35, 65); $(ED_1, ED_2)$ = (0.15, 0.35). Notation: Emp, Empirical design. | | | | | | | | |

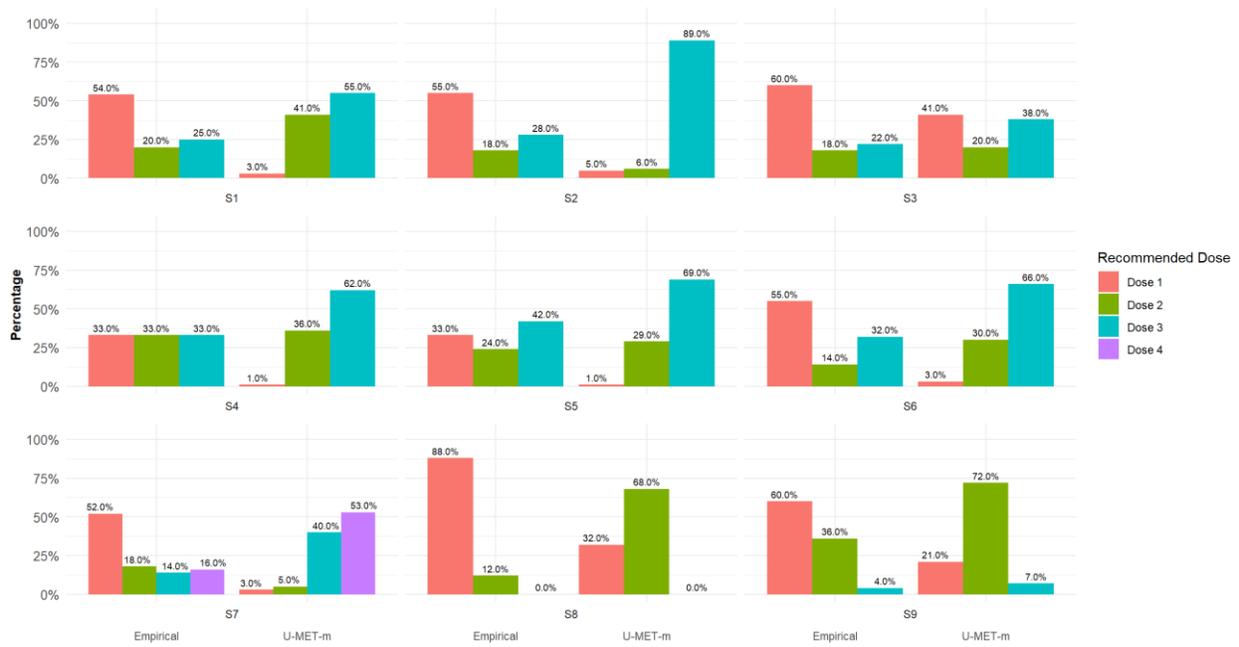

*Figure 1 Simulation results of dose selection for $n = 30$, low efficacy rate=0.23, low toxicity rate=0.13, $(u_2, u_3)$=(35, 65), ($ED_1$, $ED_2$) = (0.15, 0.35), $\alpha_1 = 0.20$, consider is low dose.*

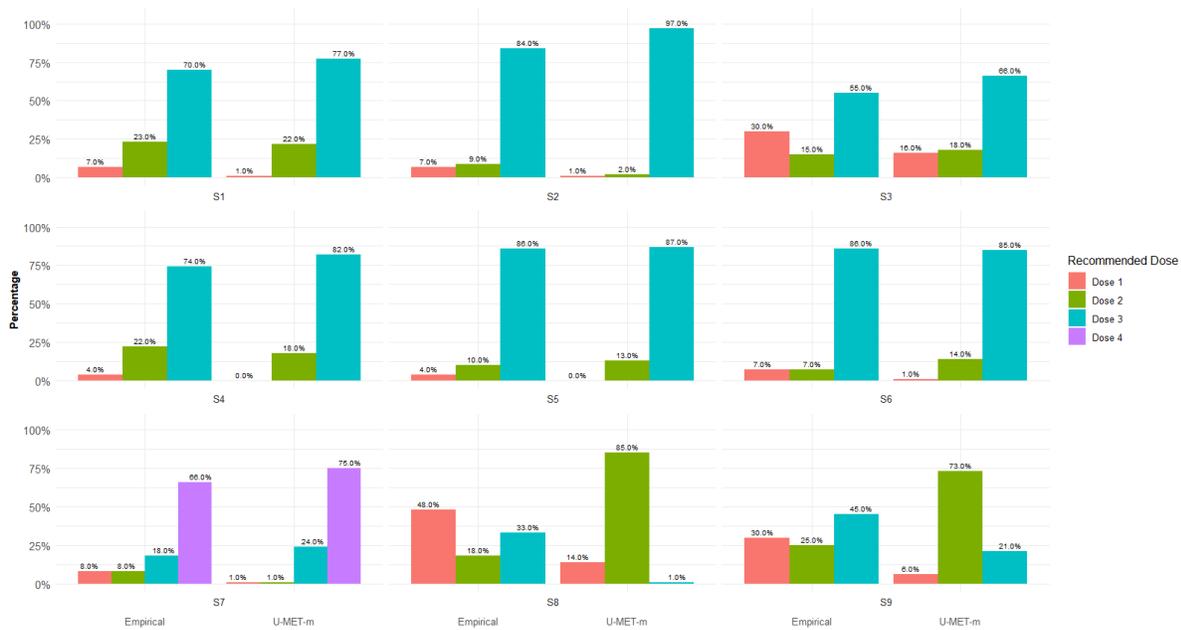

*Figure 2 Simulation results of dose selection for n=30, low efficacy rate=0.23, low toxicity rate=0.13, $(u_2, u_3)$=(35, 65), ($ED_1$, $ED_2$) = (0.15, 0.35), $\alpha_1 = 0.34$, consider is high dose.*

For the low-dose efficacy and toxicity rate of ($p_{E1} = 0.40, p_{T1} = 0.13$) (Table S2, Figure S3), when $\alpha_1 = 0.20$ and the consider zone was the low dose, the U-MET-m true dose tended to be higher than the empirical design true dose. U-MET-m performed better than the empirical approach for scenarios 2, 4, 5, 6, and 9 and about the same for scenario 1. When $\alpha_1 = 0.34$ and the consider zone was the high dose (Table S2, Figure S4), most of the time the true dose for both approaches tended to be the highest dose, except for scenarios 8 and 9. The U-MET-m was more likely than the empirical design to select its true dose across most scenarios and about the same in a few scenarios (scenarios 5 and 6).

Tables 7, S3, and S4 present the results for three endpoints. The U-MET-m and CUI-MET produced results that were almost the same and selected the same dose. For the low-dose efficacy and toxicity rate of $(p_{E1} = 0.23, p_{T1} = 0.13)$, when $\alpha_1 = 0.20$ and the consider zone was the low dose (Table 7, Figure 3), U-MET-m and CUI-MET had a higher percentage than the empirical design of selecting the correct dose for most of the scenarios (scenarios 2, 4, 5, 6, 8, and 9), and results were similar in a few scenarios (scenarios 1, 3, and 7). When $\alpha_1 = 0.34$ and the consider zone was the high dose (Table 7, Figure S5), the true dose tended to be the highest dose for most of the approaches and scenarios; exceptions were scenarios 8 and 9, in which the approaches differed in the dose selection. The U-MET-m and CUI-MET were more likely than the empirical design to select the true dose across most scenarios, and dose selections were about the same in a few scenarios (scenarios 5 and 6). Additional values of $a_1 = \{20, 32, 40\}$ were assessed to evaluate what would occur as more weight was shifted to the biomarker from the other endpoints. We were particularly interested in a value of 32 when $u_{12} = u_{13} = 50$ to assess equal weight across all endpoints of 0.34, 0.34, and 0.32 for efficacy, toxicity, and biomarker. U-MET-m and CUI-MET differed in dose selection probabilities; for some of the scenarios (scenarios 3 and 6–9), U-MET-m had an increased probability for selecting lower doses compared with CUI-MET when the weights for $u_{12}$ were increased. This difference was due to the fact that the U-MET-m mean utility decreases as $a_1$ increases, as we forced a relationship between U-MET-m and CUI-MET; in addition, as $a_1$ increases, the combined weights and overall weights are shifted. As $a_1$ increased, U-MET and CUI-MET started to differ where $u(d) < CUI(d)$, which resulted in the selection of lower

doses with U-MET-m. This difference was even more pronounced in these scenarios when the weights increased from efficacy to toxicity.

For the low-dose efficacy and toxicity rate of $(p_{E1} = 0.40, p_{T1} = 0.13)$ (Table S3, Figure S6), when $\alpha_1 = 0.20$ and the consider zone was the low dose, the U-MET-m and CUI-MET true dose tended to be higher than that produced by the empirical design. U-MET-m and CUI-MET performed better than the empirical approach for scenarios 2, 4–6, 8, and 9. When $\alpha_1 = 0.34$ and the consider zone was the high dose (Table S4, Figure S7), the true dose tended to be the highest dose for most of the scenarios and for all approaches except scenarios 8 and 9. The U-MET-m and CUI-MET were more likely than the empirical design to select its true dose across all scenarios.

| Table 7: Simulation results of percentage correct for dose selection compared to empirical design and U-MET-m truth for 3 endpoints[a] | | | | | | | | | | | |
|---|---|---|---|---|---|---|---|---|---|---|---|
| | | | | $\alpha_1$=0.2, consider is low dose | | | | $\alpha_1$=0.34, consider is high dose | | | |
| | | | | Simulation results: compare to True dose for each approach and report on % correct (list one if the same) | | | | Simulation results: compare to True dose for each approach and report on % correct (list one if the same) | | | |
| Scenario | High Efficacy | High Toxicity | High biomarker | True dose for Emp/U-MET-m/ CUI | Emp | U-MET-m | CUI-MET | True dose for Emp/U-MET-m/ CUI | Emp | U-MET-m | CUI-MET |
| 1 | (0.48,0.70) | (0.20,0.28) | (0.4,0.5) | 1/3/3 | 52/27 | 2/54 | 2/55 | 3 | 61 | 78 | 80 |
| 2 | (0.27,0.70) | (0.15,0.28) | (0.3,0.5) | 1/3/3 | 52/30 | 3/92 | 2/94 | 3 | 76 | 98 | 99 |
| 3 | (0.27,0.48) | (0.15,0.20) | (0.3,0.4) | 1/3/3 | 50/29 | 30/47 | 28/47 | 3 | 55 | 75 | 77 |
| 4 | (0.48,0.70) | (0.15,0.20) | (0.4,0.5) | 3 | 38 | 60 | 60 | 3 | 68 | 83 | 84 |
| 5 | (0.48,0.70) | (0.20, 0.20) | (0.4,0.5) | 3 | 48 | 68 | 67 | 3 | 82 | 87 | 88 |
| 6 | (0.48,0.70) | (0.28, 0.28) | (0.4,0.5) | 1/3/3 | 52/36 | 2/65 | 2/66 | 3 | 83 | 86 | 87 |
| 7 | (0.27,0.48, 0.70) | (0.15, 0.20, 0.28) | (0.3,0.4,0.5) | 1/4/4 | 54/21 | 2/54 | 1/56 | 4 | 59 | 77 | 79 |
| 8 | (0.48,0.27) | (0.15, 0.20) | (0.3,0.4) | 1/2/2 | 61/25 | 29/71 | 28/72 | 1/2/2 | 48/20 | 10/89 | 9/90 |
| 9 | (0.48,0.48) | (0.15, 0.20) | (0.3,0.4) | 1/2/2 | 50/31 | 16/77 | 15/77 | 3/2/2 | 45/26 | 21/75 | 22/75 |
| [a] Fixed parameters: $n = 30$; low dose efficacy rate, 0.23; low dose toxicity rate, 0.13; low biomarker rate, 0.2; $(u_2, u_3)$ = (35, 65); (ED$_1$, ED$_2$) = (0.15, 0.35). Notation: Emp, Empirical design. | | | | | | | | | | | |

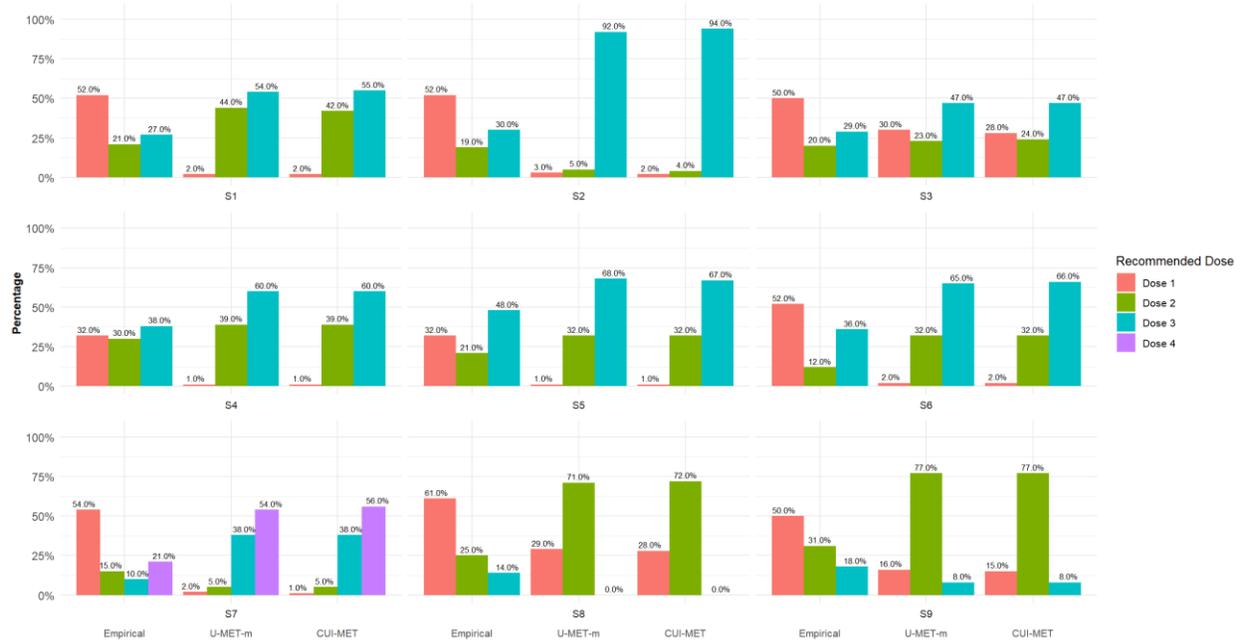

*Figure 3  Simulation results of dose selection for n=30, low efficacy rate=0.23, low toxicity rate=0.13, ($u_2$, $u_3$)=(35, 65), ($ED_1$, $ED_2$) = (0.15, 0.35), $\alpha_1$=0.20, consider is low dose.*

The Supplemental Material presents results for various utility scores [Figures S8–S15 ($u_2 = 50, u_3 = 50$) and Figures S16–S23 ($u_2 = 65, u_3 = 35$)]. As the utility score increased for $u_2$, the U-MET-m was more likely than the other approaches to select lower doses. This makes sense because as the dose search decreases, dose selection becomes more conservative, leaning toward the lower dose as it increases the dose exploration with lower $u_2$.

Overall, the U-MET-m and CUI-MET approaches performed better than the empirical design in selecting the correct dose. When $\alpha_1$ = 0.20 and the consider zone was the low dose, the empirical design tended to select the lower dose. When $\alpha_1$ = 0.34 and the consider zone was the high dose, the empirical design usually selected the same dose as U-MET-m and CUI-MET. When the utility score for $u_2$ was higher, U-MET-m tended to select the same lower dose as the empirical approach, as expected. As the third endpoint has a higher weight, U-MET-m and CUI-MET can have different dose selections, whereas the U-MET-m can lean toward a lower dose dependent on the combination and weight assignments.

# 4  Example

This study mimics an oncology phase 1 study where a dose escalation study to evaluate safety and tolerability in part 1 using the modified toxicity probability interval-2 algorithm has identified a set of safe and tolerable doses. Following this, the safety, tolerability, and antitumor activity of the investigational product were further evaluated in part 2 with a few candidate doses tested in a randomized expansion with three doses randomized. These doses need to be compared to determine the recommended phase 2 dose or the OBD based on the aggregated safety, tolerability, and efficacy data.

In the dose expansion study, each dose expansion (i.e., each dose level) has a sample size of 30 with a lower threshold of efficacy, $\phi_E = 0.35$, an upper threshold of toxicity, $\phi_T = 0.22$, and utility scores ($u_2 = 40, u_3 = 60$). There are three randomized arms dosed at 600 mg (dose 1), 1,200 mg (dose 2), or 1,800 mg (dose 3), and we wanted to determine whether the 1,800-mg dose (dose 3) increases efficacy benefit over the lower doses while maintaining an acceptable safety profile. We evaluated a sequential and a pairwise comparison. The utility scores for four combinations of efficacy and toxicity are $u_1 = 100, u_2 = 40, u_3 = 60,$ and $u_4 = 0$.

Table 8 shows two hypothetical scenarios with response and toxicity rates of the three doses, along with the dose comparison results obtained with the empirical design and the U-MET-m approaches where $\alpha_1 = 0.20$ for the sequential test. Table S5 shows the results of the pairwise comparisons using $\alpha_1 = 0.20$ and $\alpha_2 = 0.34$. The mean utility difference is $u_\Delta = u_2(d) - u_1(d)$.

Suppose we have additional data available and add a third endpoint, circulating tumor DNA, to demonstrate three endpoints and the results produced from the additional CUI-MET approach. We used the first scenario described in section 2.4, in which U-MET-m is related to CUI-MET where $a_1 = a_4 = 20$, $a_2 = a_3 = 10$, and utility scores are ($u_{11} = u_{12} + u_{13}, u_{12} = 40, u_{13} = 60, u_{14} = 0, u_{01} = u_{02} +$ $u_{03}, u_{02} = 30, u_{03} = 50$, and $u_{04} = 0$). The weight of $a_1 = 20$ produces a weight for the biomarker of 0.20 for CUI-MET. Table 9 reports the response, toxicity, and biomarker rates of the three doses. The dose comparison results using the empirical design, U-MET-m, and CUI-MET approaches are presented with $\alpha_1 = 0.20$ for the sequential test (Table 9).

| Table 8 Results of dose comparison from examples for sequential test | | | | |
|---|---|---|---|---|
| | Response (%) | Toxicity (%) | Empirical design | U-MET-m |
| Scenario | | | ED, TR, dose | $\hat{u}_\Delta^*, \Pr(\hat{u}_\Delta^* > 0)$, dose |
| 1 (Dose 1, 2, 3) | (47, 57, 76) | (17, 20, 26) | | |
| 1, step 1 (Dose 1 vs 3) | 47 vs 76 | 17 vs 26 | 29%, 1.53, 1* | 13.8, 0.870, 3 |
| 1, step 2 (Dose 2 vs 3) | 57 vs 76 | 20 vs 26 | Stops at dose 1, no further comparison | 9, 0.773, 2* |
| 2 (Dose 1, 2, 3) | (47, 67, 60) | (17, 20, 26) | | |
| 2, step 1 (Dose 1 vs 3) | 47 vs 60 | 17 vs 26 | 13%, 1.53, 1* | Removes dose 3 from comparison |
| 2, step 2 (Dose 1 vs 2) | 47 vs 67 | 20 vs 26 | Stops at dose 1, no further comparison | 10.80, 0.808, 2* |
| * Dose selected | | | | |

The U-MET-m and CUI-MET approaches tended to produce results that were similar to and discordant with the empirical design. When there were two endpoints, U-MET-m favored the higher dose and the empirical design stopped at the lowest dose for both scenarios (Tables 8 and S5). The sequential test and the pairwise test showed similar findings for U-MET-m, where the second dose was the best dose for both scenarios. The empirical design did not arrive at a similar conclusion and selected the lowest dose with the sequential test; however, when the pairwise comparison allowed the consider zone, the empirical design leaned toward the high dose for scenario 1 and the lowest dose for scenario 2. When accounting for a third endpoint, the U-MET-m and CUI-MET tended to agree and differed from the empirical design (Table 9). In the final step of scenario 1, the U-MET-m and CUI-MET approaches came to different conclusions, whereas U-MET-m selected dose 2 and CUI-MET selected dose 3; however, the probabilities were borderline and quite close 0.80. In scenario 1, the empirical design selected dose 1. Although in scenarios 2 and 3 both U-MET-m and CUI-MET agreed and selected dose 2, the empirical design selected the lowest dose for all scenarios.

| Table 9 Results of dose comparison from examples for sequential test | | | | | | |
|---|---|---|---|---|---|---|
| | Response (%) | Toxicity (%) | Biomarker (%) | Empirical design | U-MET-m | CUI-MET |
| Scenario | | | | ED, TR, BD, dose | $\hat{u}_\Delta^*$, $\Pr(\hat{u}_\Delta^* > 0)$, dose | $CUI_\Delta^*$, $\Pr(\widehat{CUI}_\Delta^* > 0)$, dose |
| 1 (Dose 1, 2, 3) | (47, 57, 76) | (17, 20, 26) | (25, 30, 45) | | | |
| 1, step 1 (Dose 1 vs 3) | 47 vs 76 | 17 vs 26 | 25 vs 45 | 29%, 1.53, 0.2,1* | 15.3, 0.882, 3 | 15.8, 0.892, 3 |
| 1, step 2 (Dose 2 vs 3) | 57 vs 76 | 20 vs 26 | 30 vs 45 | Stops at dose 1, no further comparison | 10.3, 0.791, 2* | 10.7, 0.801, 3* |
| 2 (Dose 1, 2, 3) | (47, 57, 76) | (17, 20, 26) | (25, 40, 35) | | | |
| 2, step 1 (Dose 1 vs 3) | 47 vs 76 | 17 vs 26 | 25 vs 35 | 29%, 1.53, 0.1, 1* | 13.8, 0.857, 3 | 13.8, 0.858, 3 |
| 2, step 2 (Dose 2 vs 3) | 57 vs 76 | 20 vs 26 | 40 vs 35 | Stops at dose 1, no further comparison | 7.5, 0.720, 2* | 6.7, 0.702, 2* |
| 3 (Dose 1, 2, 3) | (47, 67, 60) | (17, 20, 26) | (25, 45, 45) | | | |
| 3, step 1 (Dose 1 vs 3) | 47 vs 60 | 17 vs 26 | 25 vs 45 | 13%, 1.5, 0.2, 1* | Removes dose 3 from comparison | Removes dose 3 from comparison |
| 3, step 2 (Dose 1 vs 2) | 47 vs 67 | 17 vs 20 | 25 vs 45 | Stops at dose 1, no further comparison | 12.5, 0.832, 2* | 13.1, 0.846, 2* |
| * Dose selected | | | | | | |

These hypothetical trial results demonstrate how decisions are made with the empirical design versus the U-MET-m and CUI-MET approaches. Both the U-MET-m and CUI-MET approaches led to the same dose decision most of the time, whereas U-MET-m assigned weights jointly and CUI-MET assigned weights marginally. Even so, the final utility values were similar for U-MET-m and CUI-MET. The U-MET-m and CUI-MET approaches differed from the empirical design, which was more conservative in dose selection. Based on these observations, the U-MET-m and CUI-MET approaches tended to select a higher dose and a clear decision where these approaches account for uncertainty in the data and are based on a degree of evidence. In deciding between U-MET-m and CUI-MET, it is important to note whether joint weights are to be accounted for to select the appropriate approach. At the very least, it would be advisable to run the U-MET-m or CUI-MET approach as additional support of the empirical design to help make a decision for a final dose.

# 5 Discussion

The advent of dose optimization in oncology has led to a shift in the approaches used to determine the OBD and MTD. Randomized, dosed-arm, phase 2 trials can facilitate additional investigation of the identified OBD and MTD in a targeted population by incorporating safety, efficacy, and biomarker data. With the newer concept of randomized, dosed-arm, phase 1/2 trials in oncology, more guidance is needed on how to select the optimal dose, especially when there are multiple doses and endpoints.

We propose the U-MET-m and CUI-MET designs, which are based on the U-MET framework, to identify the optimal dose in randomized dose expansion trials. Here we demonstrate the use of these approaches when there are at least two doses that need to be compared with two to three endpoints. Although the U-MET-m and CUI-MET approaches are similar, a major difference is that U-MET-m accounts for multiple endpoints jointly, whereas CUI-MET accounts for endpoints marginally. If more than three endpoints are to be included, CUI-MET may be simpler to decide on weights, although the operating characteristics need to be evaluated when there are more than three endpoints. The clinical team may decide whether they want to assign weights jointly or marginally.

We performed simulation studies and a hypothetical oncology study to demonstrate and compare the empirical decision table to the U-MET-m and CUI-MET approaches. Numerical studies demonstrated that, compared with the empirical approach, the U-MET-m and CUI-MET designs performed better at selecting both higher and lower doses. When $u_3 \geq u_2$, it is implied that patients can tolerate higher toxicity for higher efficacy, leading to a more comprehensive exploration of the dose space. We showed similar operating characteristics across both approaches. We noticed that the CUI value can be slightly higher than the U-MET-m value, which in borderline cases could result in a different decision. The advantage of the CUI-MET is the simplicity and ease of combining multiple endpoints, with the caveat that weights need to be assigned for the contribution of each endpoint. Compared with the empirical approach, the U-MET-m and CUI-MET approaches offer several additional advantages. These approaches account for uncertainty in the data by using posterior probabilities, allow a flexible approach to customize the risk-benefit tradeoff, and provide a theoretical decision framework to measure the evidence on dose selection.

This manuscript demonstrated using U-MET-m and CUI-MET to address comparing randomized dosed arms in an expansion trial with at least two doses and multiple endpoints. We recommend the use of the U-MET-m and CUI-MET approaches to compare randomized dosed arms or to provide supportive evidence for an empirical decision table. Future work needs to be performed on conditional and marginal probabilities and on weight assignments to understand the impact on CUI-MET-m and U-MET-m. We assumed the endpoints to be independent and will address correlation in future work. We will also address delays in efficacy and sensitivity analysis to missing data. There is added value to dose optimization with the utility comparison approaches in dose expansion trials with randomized dosed arms. Software will be available and the utility approaches will be implemented and packed into an Rshiny App.

**Acknowledgments**
We thank Deborah Shuman of AstraZeneca for editorial support.

**Declaration of Interests**
G.D. and D.R. are employees of AstraZeneca and may have stock ownership, options, and/or interests in the company.

**Funding**
This study was funded by AstraZeneca.

**References**

Coffey, T., Gennings, C., and Moser, V. C. (2007), "The simultaneous analysis of discrete and continuous outcomes in a dose-response study: using desirability functions," *Regul Toxicol Pharmacol*, 48 (1), 51-58. DOI: 10.1016/j.yrtph.2006.12.004.

D'Angelo, G., Gong, M., Marshall, J., Yuan, Y., and Li, X. (2024), "U-MET: Utility-Based Dose Optimization Approach for Multiple-Dose Randomized Trial Designs," *Statistics in Biopharmaceutical Research*, 1–11. DOI: DOI: 10.1080/19466315.2024.2365630.

FDA, U. S. 2023. Project optimus. Optimizing the Dosage of Human Prescription Drugs and Biological Products for the Treatment of Oncologic Diseases [Internet].

Guo, B., and Yuan, Y. (2023), "DROID: dose-ranging approach to optimizing dose in oncology drug development," *Biometrics*. DOI: 10.1111/biom.13840.

Jiang, Z., Mi, G., Lin, J., Lorenzato, C., and Ji, Y. (2023), "A Multi-Arm Two-Stage (MATS) design for proof-of-concept and dose optimization in early-phase oncology trials," *Contemp Clin Trials*, 132, 107278. DOI: 10.1016/j.cct.2023.107278.

Li, Y., Zhang, Y., Mi, G., and Lin, J. (2024), "A seamless phase II/III design with dose optimization for oncology drug development," *Stat Med*, 43 (18), 3383-3402. DOI: 10.1002/sim.10129.

Lin, R., Zhou, Y., Yan, F., Li, D., and Yuan, Y. (2020), "BOIN12: Bayesian Optimal Interval Phase I/II Trial Design for Utility-Based Dose Finding in Immunotherapy and Targeted Therapies," *JCO Precis Oncol*, 4. DOI: 10.1200/PO.20.00257.

Ouellet, D. (2010), "Benefit-risk assessment: the use of clinical utility index," *Expert Opin Drug Saf*, 9 (2), 289-300. DOI: 10.1517/14740330903499265.

Ouellet, D., Werth, J., Parekh, N., Feltner, D., McCarthy, B., and Lalonde, R. L. (2009), "The use of a clinical utility index to compare insomnia compounds: a quantitative basis for benefit-risk assessment," *Clin Pharmacol Ther*, 85 (3), 277-282. DOI: 10.1038/clpt.2008.235.

Song, M. K., Lin, F. C., Ward, S. E., and Fine, J. P. (2013), "Composite variables: when and how," *Nurs Res*, 62 (1), 45-49. DOI: 10.1097/NNR.0b013e3182741948.

Sverdlov, O., Ryeznik, Y., and Wu, S. (2015), "Exact Bayesian Inference Comparing Binomial Proportions, With Application to Proof-of-Concept Clinical Trials," *Ther Innov Regul Sci*, 49 (1), 163-174. DOI: 10.1177/2168479014547420.

Winzenborg, I., Soliman, A. M., and Shebley, M. (2021), "A Personalized Medicine Approach Using Clinical Utility Index and Exposure-Response Modeling Informed by Patient Preferences Data," *CPT Pharmacometrics Syst Pharmacol*, 10 (1), 40-47. DOI: 10.1002/psp4.12570.

Yang, P., Li, D., Lin, R., Huang, B., and Yuan, Y. (2024), "Design and sample size determination for multiple-dose randomized phase II trials for dose optimization," *Stat Med*, 43 (15), 2972-2986. DOI: 10.1002/sim.10093.

Zhou, Y., Lee, J. J., and Yuan, Y. (2019), "A utility-based Bayesian optimal interval (U-BOIN) phase I/II design to identify the optimal biological dose for targeted and immune therapies," *Stat Med*, 38 (28), 5299-5316. DOI: 10.1002/sim.8361.

# Utility-based Dose Optimization Approaches for Multiple-dose Randomized Trial Designs accounting for multiple endpoints

Gina D'Angelo, Guannan Chen, Di Ran

Supplemental Material

## Simulation data generation details

When there were two endpoints, such as toxicity and efficacy, we generated the data as follows. For each patient at each dose level, we simulated a bivariate normal random variable ($z_{Td}$, $z_{Ed}$) with zero mean vector and the covariance matrix of $\begin{bmatrix} 1 & \rho_{TEd} \\ \rho_{TEd} & 1 \end{bmatrix}$, where $\rho_{TEd}$ is the toxicity and efficacy correlation for the $dth$ dose, where $d = 1, \dots, D$ for lowest to highest dose. The outcomes can be obtained as

$$\text{Toxicity outcome: } x_{Td} = I\{z_{Td} \leq \Phi^{-1}(p_T(d))\}$$

$$\text{Efficacy outcome: } x_{Ed} = I\{z_{Ed} \leq \Phi^{-1}(p_E(d))\},$$

where $p_T(d)$ and $p_E(d)$ are the marginal rates for toxicity and efficacy, $\Phi^{-1}$ is the inverse of the standard normal cumulative distribution function, $d = 1, \dots, D$ for lowest to highest dose, and $I$ is the indicator function. We assumed $\phi_T = 0.35$, $\phi_E = 0.22$, $c_T = 0.95$, and $c_E = 0.9$ in our simulations and $\rho_{TE} = 0$. For our purposes, we generated the data to have no correlations, and in previous evaluations we found correlation to not have an impact; however, this can be explored in future work.

When there are three endpoints, such as efficacy, toxicity, and a biomarker, we simulated the data as follows. For each patient at each dose level, we simulated a multivariate normal random variable ($z_{Td}$, $z_{Ed}$, $z_{Bd}$) with zero mean vector and the covariance matrix of

$$\begin{bmatrix} 1 & \rho_{TEd} & \rho_{TBd} \\ \rho_{TEd} & 1 & \rho_{EBd} \\ \rho_{TBd} & \rho_{EBd} & 1 \end{bmatrix},$$

where $\rho_{TEd}$ is the toxicity and efficacy correlation, $\rho_{TBd}$ is the toxicity and biomarker correlation, and $\rho_{EBd}$ is the efficacy and biomarker correlation for the $dth$ dose. The outcomes can be obtained as

$$\text{Toxicity outcome: } x_{Td} = I\{z_{Td} \leq \Phi^{-1}(p_T(d))\}$$

$$\text{Efficacy outcome: } x_{Ed} = I\{z_{Ed} \leq \Phi^{-1}(p_E(d))\}$$

$$\text{Biomarker outcome: } x_{Bd} = I\{z_Bd \leq \Phi^{-1}(p_B(d))\},$$

where $p_T(d)$, $p_E(d)$, and $p_B(d)$ are the marginal rates for toxicity, efficacy, and biomarker, respectively; $\Phi^{-1}$ is the inverse of the standard normal cumulative distribution function; $d = 1, ..., D$ for lowest to highest dose; and $I$ is the indicator function. We assumed $\rho_{TEd} = 0$, $\rho_{TBd} = 0$, and $\rho_{EBd} = 0$. We set correlation to be 0.

## Simulation results

Table S1 shows the specified cut-offs for the empirical design used in the simulation studies and the example.

| Table S1: Empirical design example | | | |
|---|---|---|---|
| Toxicity Ratio (Tox$_H$/Tox$_L$) | **Efficacy difference (Eff$_H$– Eff$_L$)** | | |
| | >35% | [15,35%] | <15% |
| <1.5 | High | High | Consider* |
| [1.5,2] | High | Consider | Low |
| >2x | Consider | Low | Low |
| * An alternative dose selection for the scenario of ED<0 can be Low dose, and this will be indicated a priori. Abbreviations: Eff$_H$, Eff$_L$ = efficacy of the higher and lower doses, respectively; Tox$_H$, Tox$_L$ = toxicity of the higher and lower doses, respectively. | | | |

Results from the first simulation set-up in Table 5 are shown in Figure S1. Additional analyses for the first simulation set-up were run using different decision criteria of $\alpha_1 = 0.34$ and consider is high dose; results are reported in Figure S2.

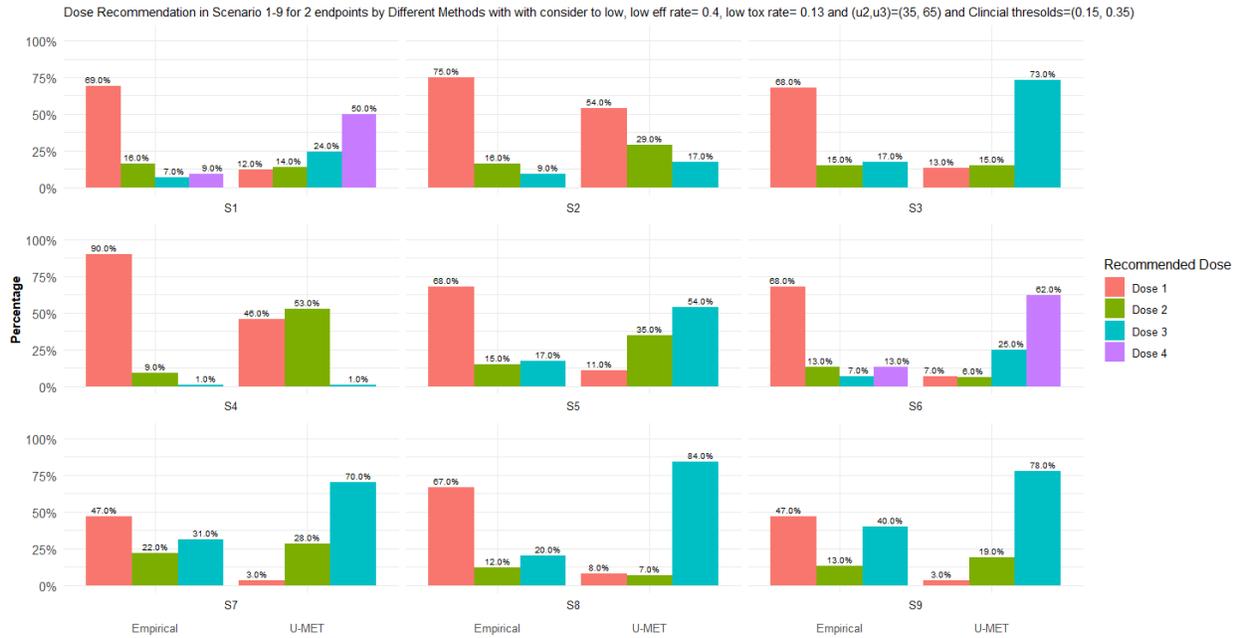

*Figure S1 Simulation results of dose selection for n=40, low efficacy rate=0.23, low toxicity rate=0.13, ($u_2$, $u_3$)=(35, 65), ($ED_1$, $ED_2$) = (0.15, 0.35), $\alpha_1 = 0.20$, consider is low dose.*

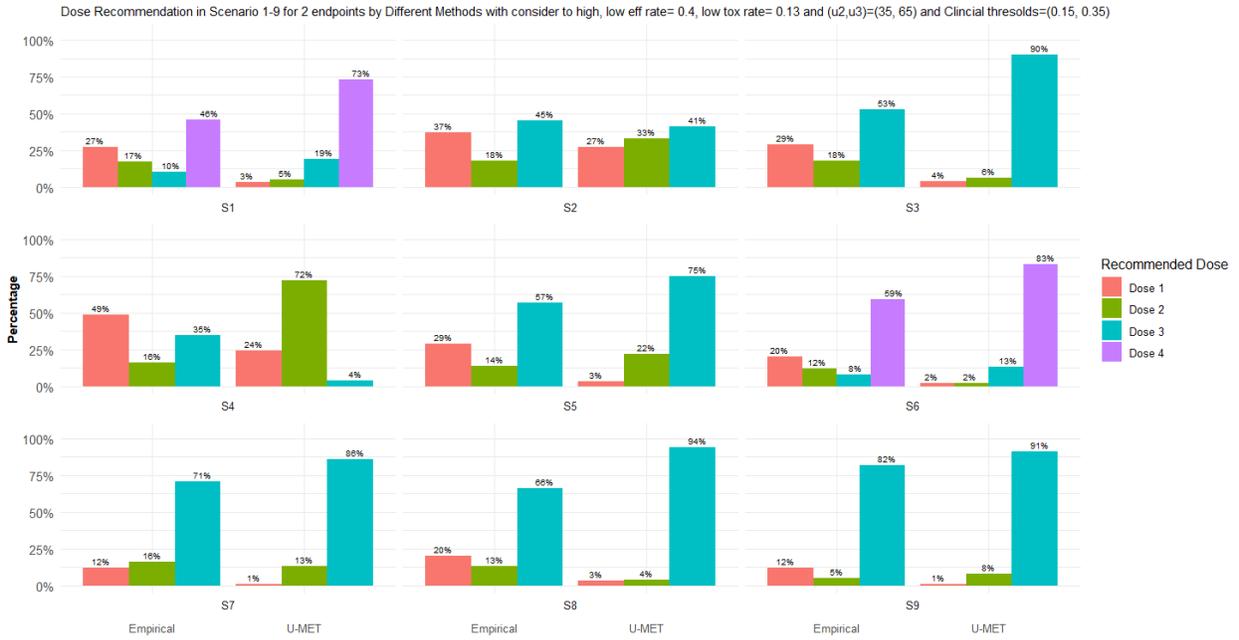

*Figure S2 Simulation results of dose selection for n=40, low efficacy rate=0.23, $(u_2, u_3)$=(35, 65), $(ED_1, ED_2)$ = (0.15, 0.35), $\alpha_1 = 0.34$, consider is high dose.*

Table S2 presents the results supporting Figures S3 and S4 of two endpoints with the low-dose efficacy and toxicity rate of $(p_{E1} = 0.40, p_{T1} = 0.13)$.

| Table S2: Simulation results of percentage correct for dose selection compared to empirical design and U-MET-m truth[a] | | | | | | | | |
|---|---|---|---|---|---|---|---|---|
| | | | $\alpha_1$=0.2, consider is low dose | | | $\alpha_1$=0.34, consider is high dose | | |
| | | | Simulation results: compare to True dose for each approach and report on % correct (only list one if the same) | | | Simulation results: compare to True dose for each approach and report on % correct (only list one if the same) | | |
| Scenario | High Eff | High Tox | True dose for Empirical/U-MET-m (only list one if the same) | Empirical decision | U-MET-m | True dose for Empirical /U-MET-m (only list one if the same) | Empirical decision | U-MET-m |
| S1 | (0.65,0.85) | (0.20,0.28) | 1/2 | 58/22 | 3/52 | 3 | 65 | 76 |
| S2 | (0.47,0.85) | (0.15,0.28) | 1/3 | 58/23 | 4/87 | 3 | 75 | 95 |
| S3 | (0.47,0.65) | (0.15,0.20) | 1/2 | 61/20 | 32/39 | 3 | 55 | 65 |
| S4 | (0.65,0.85) | (0.15,0.20) | 3 | 29 | 59 | 3 | 69 | 81 |
| S5 | (0.65,0.85) | (0.20, 0.20) | 3 | 37 | 67 | 3 | 83 | 86 |
| S6 | (0.65,0.85) | (0.28, 0.28) | 1/3 | 58/26 | 3/65 | 3 | 83 | 84 |
| S7 | (0.47,0.65, 0.85) | (0.15, 0.20, 0.28) | 1/3 | 58/13 | 4/39 | 4 | 58 | 74 |
| S8 | (0.65,0.47) | (0.15, 0.20) | 1/2 | 86/13 | 28/76 | 1/2 | 50/17 | 11/88 |
| S9 | (0.65, 0.65) | (0.15, 0.20) | 1/2 | 61/36 | 17/76 | 3/2 | 44/27 | 22/73 |
| [a] Fixed parameters: $n = 30$; low dose efficacy rate, 0.40; low dose toxicity rate, 0.13; $(u_2, u_3)$ = (35, 65); (ED$_1$, ED$_2$) = (0.15, 0.35). | | | | | | | | |

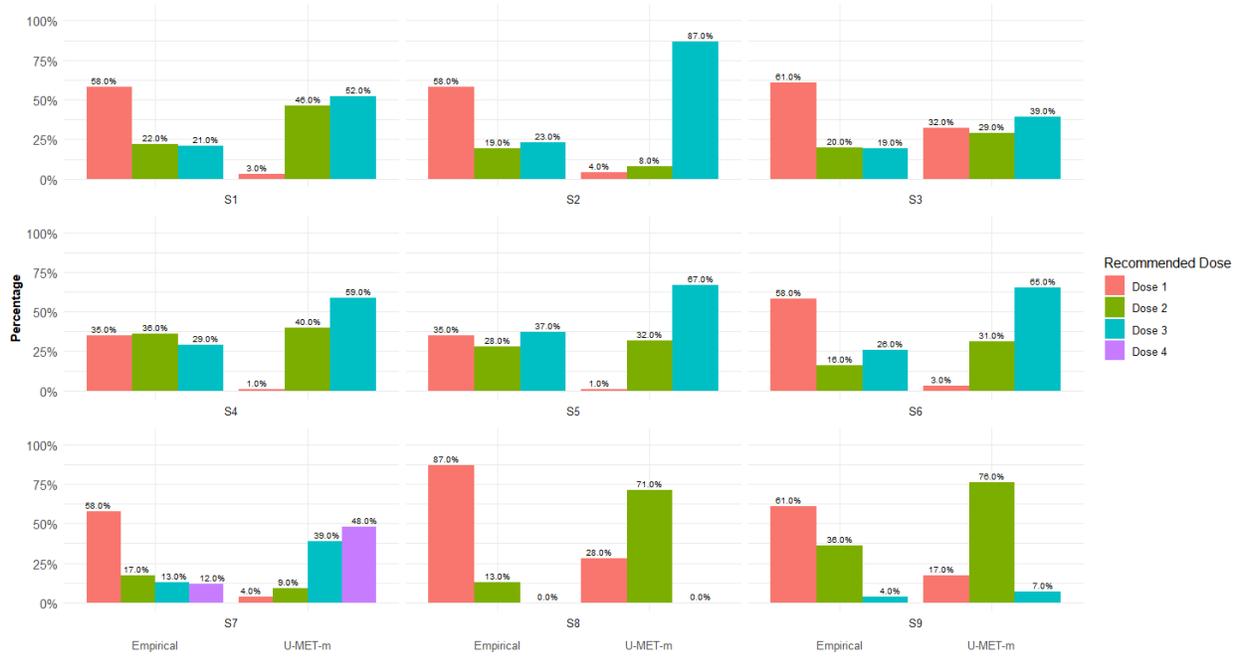

*Figure S3 Simulation results of dose selection for n=30, low efficacy rate=0.40, low toxicity rate=0.13, $(u_2, u_3)$=(35, 65), $(ED_1, ED_2)$ = (0.15, 0.35), $\alpha_1 = 0.20$, consider is low dose (Table S2).*

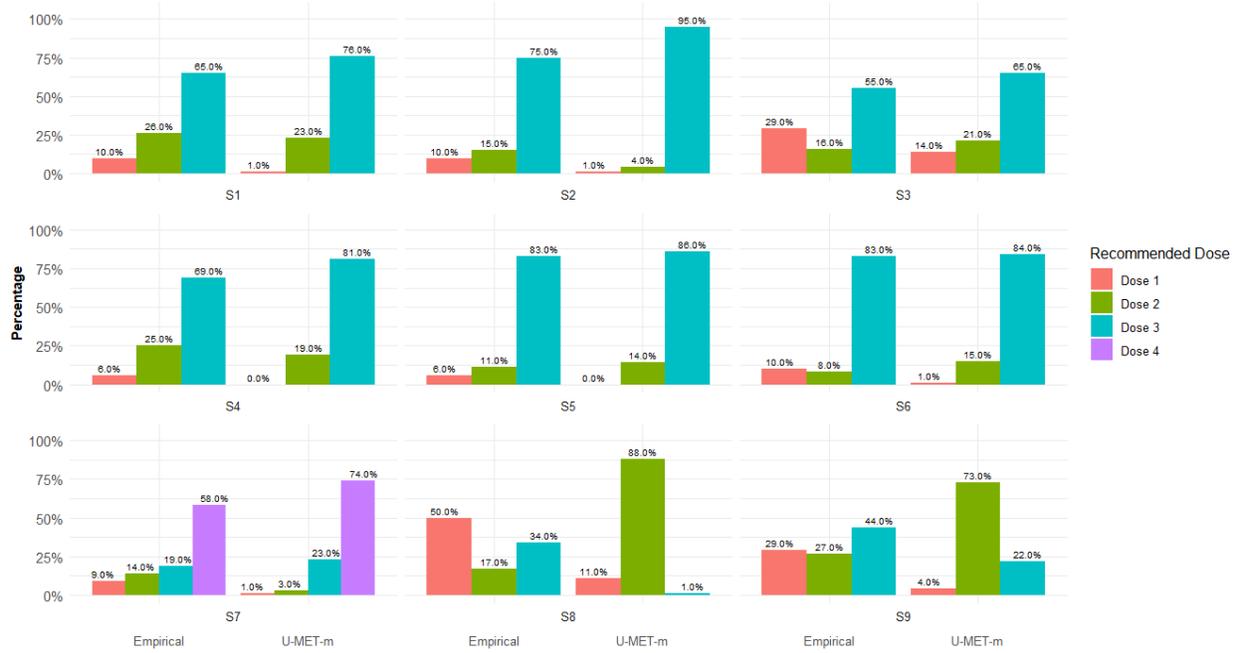

*Figure S4 Simulation results of dose selection for n=30, low efficacy rate=0.40, low toxicity rate=0.13, $(u_2, u_3)$=(35, 65), $(ED_1, ED_2)$ = (0.15, 0.35), $\alpha_1 = 0.34$, consider is high dose (Table S2).*

Figure S5 presents the scenario of three endpoints with low-dose efficacy and toxicity rates of $(p_{E1} = 0.23, p_{T1} = 0.13)$, $\alpha_1 = 0.34$ and the consider zone is select high dose. The supporting table and figure of three endpoints with the low-dose efficacy and toxicity rate of $(p_{E1} = 0.40, p_{T1} = 0.13)$, $\alpha_1 = 0.20$ and the consider zone is select the low dose are shown in Table S3 and Figure S6.

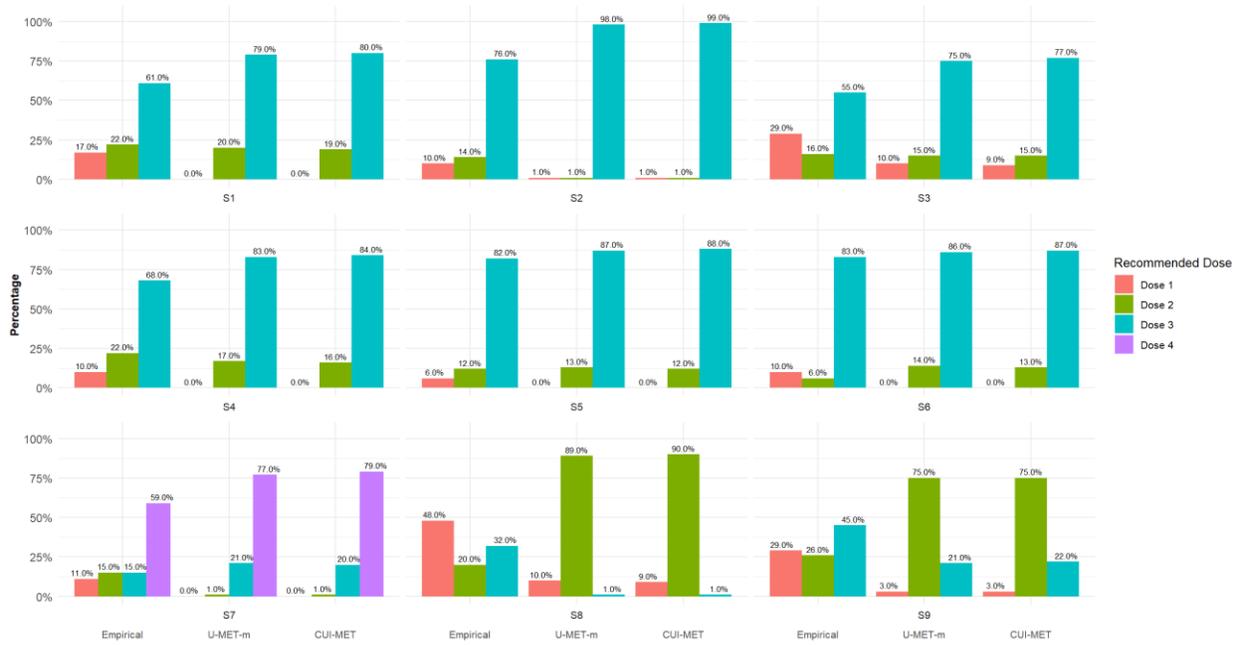

*Figure S5 Simulation results of dose selection for n=30, low efficacy rate=0.23, low toxicity rate=0.13, $(u_2, u_3)$=(35, 65), $(ED_1, ED_2)$ = (0.15, 0.35), $\alpha_1$=0.34, consider is high dose (Table 7).*

| Table S3: Simulation results of percentage correct for dose selection compared to empirical design and U-MET-m truth for 3 endpoints [a] | | | | | | | |
|---|---|---|---|---|---|---|---|
| Compare all doses | | | | | Simulation results: compare to True dose for each approach and report on % correct (only list one if the same) | | |
| | | | | | Empirical decision | U-MET-m | CUI-MET |
| Scenario | High Eff | High Tox | High biomarker | True dose selection for Empirical/U-MET-m/CUI (only list one if the same) | | | |
| S1 | (0.65,0.85) | (0.20,0.28) | (0.4,0.5) | 1/2/2 | 55/20 | 3/45 | 3/44 |
| S2 | (0.47,0.85) | (0.15,0.28) | (0.3,0.5) | 1/3/3 | 55/26 | 3/90 | 2/91 |
| S3 | (0.47,0.65) | (0.15,0.20) | (0.3,0.4) | 1/2/2 | 52/22 | 29/32 | 27/33 |
| S4 | (0.65,0.85) | (0.15,0.20) | (0.4,0.5) | 3 | 36 | 59 | 60 |
| S5 | (0.65,0.85) | (0.20, 0.20) | (0.4,0.5) | 3 | 46 | 67 | 68 |
| S6 | (0.65,0.85) | (0.28, 0.28) | (0.4,0.5) | 1/3/3 | 55/33 | 2/65 | 2/66 |
| S7 | (0.47,0.65,0.85) | (0.15, 0.20, 0.28) | (0.3,0.4,0.5) | 1/3/3 | 58/9 | 2/37 | 2/38 |
| S8 | (0.65,0.47) | (0.15, 0.20) | (0.3,0.4) | 1/2/2 | 64/22 | 28/72 | 27/73 |
| S9 | (0.65, 0.65) | (0.15, 0.20) | (0.3,0.4) | 1/2/2 | 52/30 | 16/74 | 15/75 |
| [a] Fixed parameters: $n = 30$; low dose efficacy rate, 0.40; low dose toxicity rate, 0.13; low biomarker rate, 0.2; $(u_2, u_3) = (35, 65)$; $(ED_1, ED_2) = (0.15, 0.35)$; $\alpha_1 = 0.20$, consider is low dose | | | | | | | |

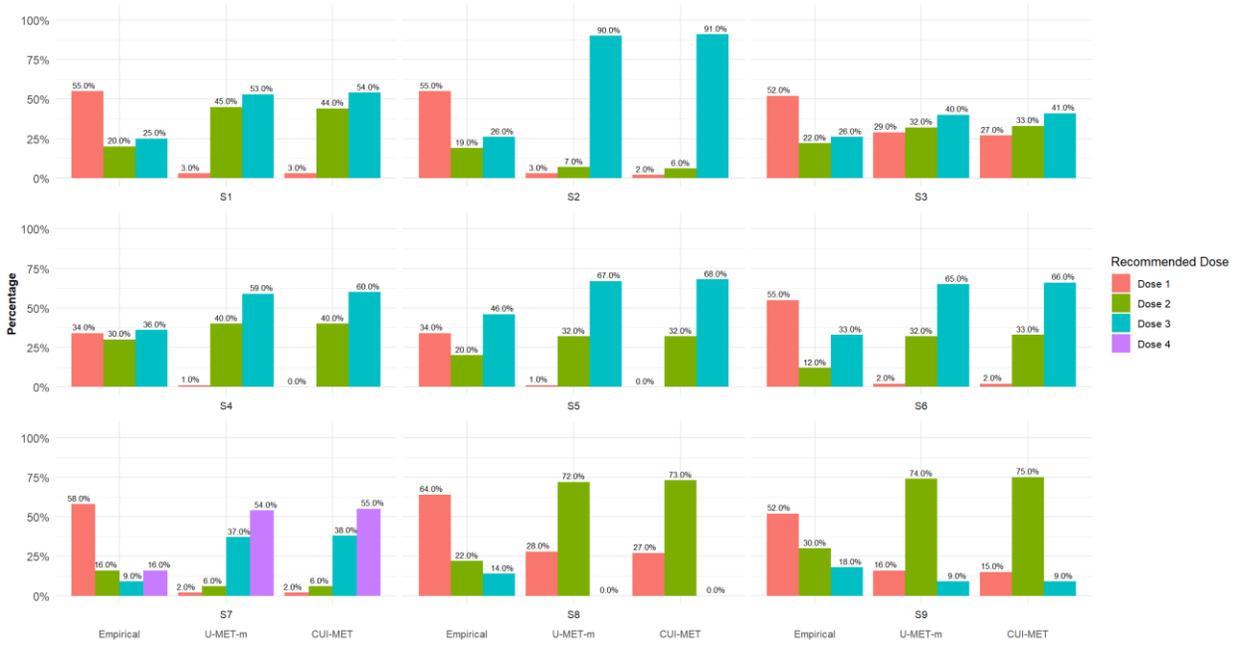

Figure S6 Simulation results of dose selection for n=30, low efficacy rate=0.40, low toxicity rate=0.13, $(u_2, u_3)=(35, 65)$, $(ED_1, ED_2) = (0.15, 0.35)$, $\alpha_1=0.20$, consider is low dose.

Table S4 and Figure S7 present the scenario of three endpoints with the low-dose efficacy and toxicity rate of $(p_{E1} = 0.40, p_{T1} = 0.13)$, $\alpha_1 = 0.34$, and the consider zone is select the high dose.

| Table S4: Simulation results of percentage correct for dose selection compared to empirical design and U-MET-m truth for 3 endpoints [a] | | | | | | | |
|---|---|---|---|---|---|---|---|
| Compare all doses | | | | Simulation results: compare to True dose for each approach and report on % correct (only list one if the same) | | | |
| | | | | | Empirical decision | U-MET-m | CUI-MET |
| Scenario | High Eff | High Tox | High biomarker | True dose selection for Empirical/U-MET-m/CUI (only list one if the same) | | | |
| S1 | (0.65,0.85) | (0.20,0.28) | (0.4,0.5) | 3 | 59 | 78 | 79 |
| S2 | (0.47,0.85) | (0.15,0.28) | (0.3,0.5) | 3 | 69 | 98 | 98 |
| S3 | (0.47,0.65) | (0.15,0.20) | (0.3,0.4) | 3 | 54 | 69 | 69 |
| S4 | (0.65,0.85) | (0.15,0.20) | (0.4,0.5) | 3 | 66 | 83 | 83 |
| S5 | (0.65,0.85) | (0.20, 0.20) | (0.4,0.5) | 3 | 80 | 87 | 88 |
| S6 | (0.65,0.85) | (0.28, 0.28) | (0.4,0.5) | 3 | 80 | 87 | 87 |
| S7 | (0.47,0.65,0.85) | (0.15, 0.20, 0.28) | (0.3,0.4,0.5) | 4 | 53 | 78 | 79 |
| S8 | (0.65,0.47) | (0.15, 0.20) | (0.3,0.4) | 1/2/2 | 47/18 | 10/88 | 9/89 |
| S9 | (0.65, 0.65) | (0.15, 0.20) | (0.3,0.4) | 3/2/2 | 45/26 | 23/73 | 23/72 |
| [a] Fixed parameters: $n = 30$; low dose efficacy rate, 0.40; low dose toxicity rate, 0.13; low biomarker rate, 0.2; $(u_2, u_3)$ = (35, 65); (ED$_1$, ED$_2$) = (0.15, 0.35); $\alpha_1$=0.34, consider is high dose | | | | | | | |

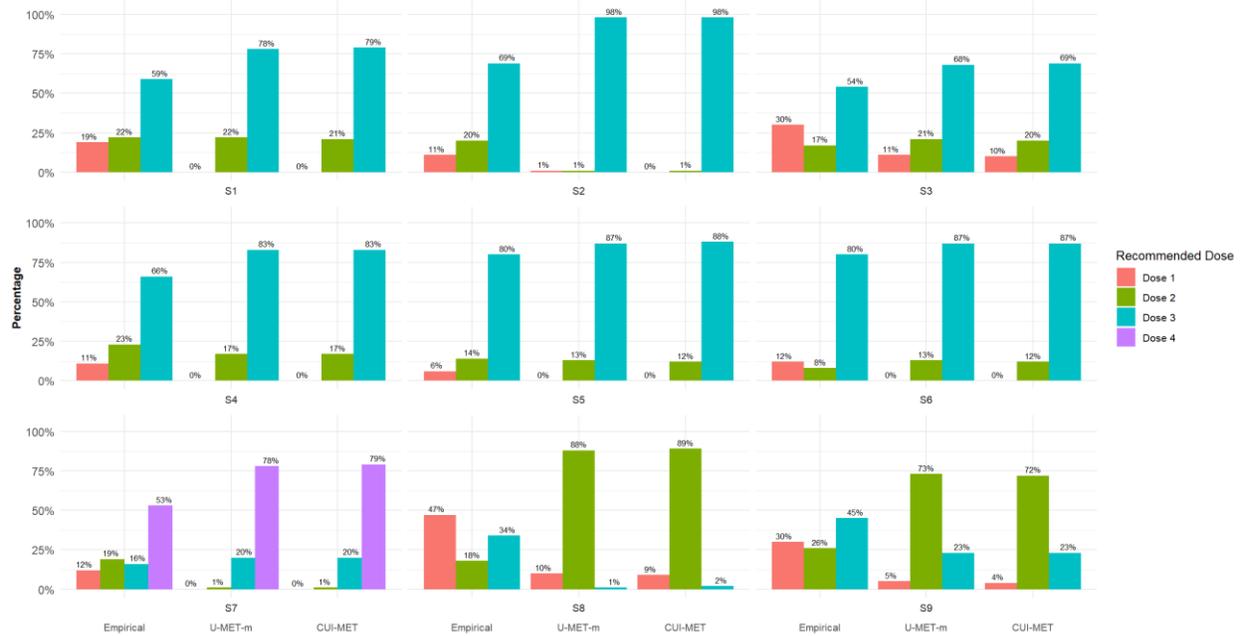

*Figure S7 Simulation results of dose selection for n=30, low efficacy rate=0.40, low toxicity rate=0.13, $(u_2, u_3)$=(35, 65), (ED$_1$, ED$_2$) = (0.15, 0.35), $\alpha_1$=0.34, consider is high dose.*

Additional analyses for the second simulation setup were run using different utility scores for Table 1, where ($u_2 = 50, u_3 = 50$). The utility score was changed to put equal weight on $u_2$ and $u_3$ to demonstrate what happens when there is the same weight for efficacy and toxicity as no efficacy and no toxicity. Results are shown in Figures S8–S15.

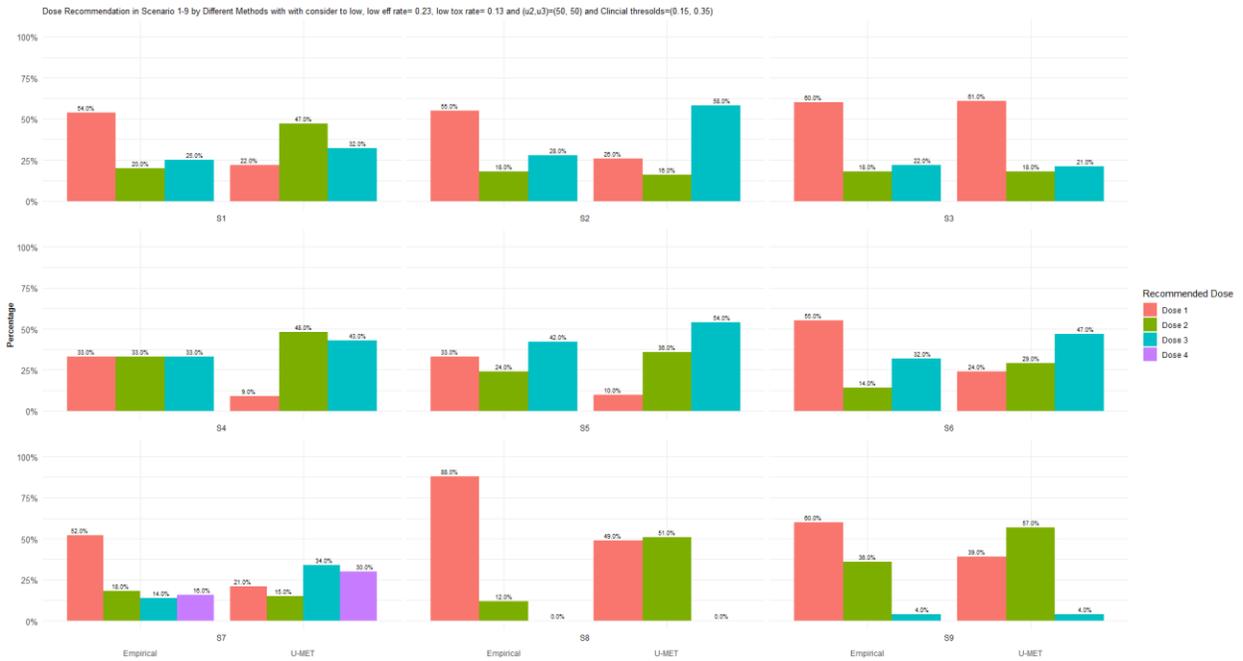

*Figure S8 Simulation results of dose selection for n=30, low efficacy rate=0.23, low toxicity rate=0.13, ($u_2$, $u_3$)=(50, 50), ($ED_1$, $ED_2$) = (0.15, 0.35), $\alpha_1 = 0.20$, consider is low dose (Same set-up as Table 6)*

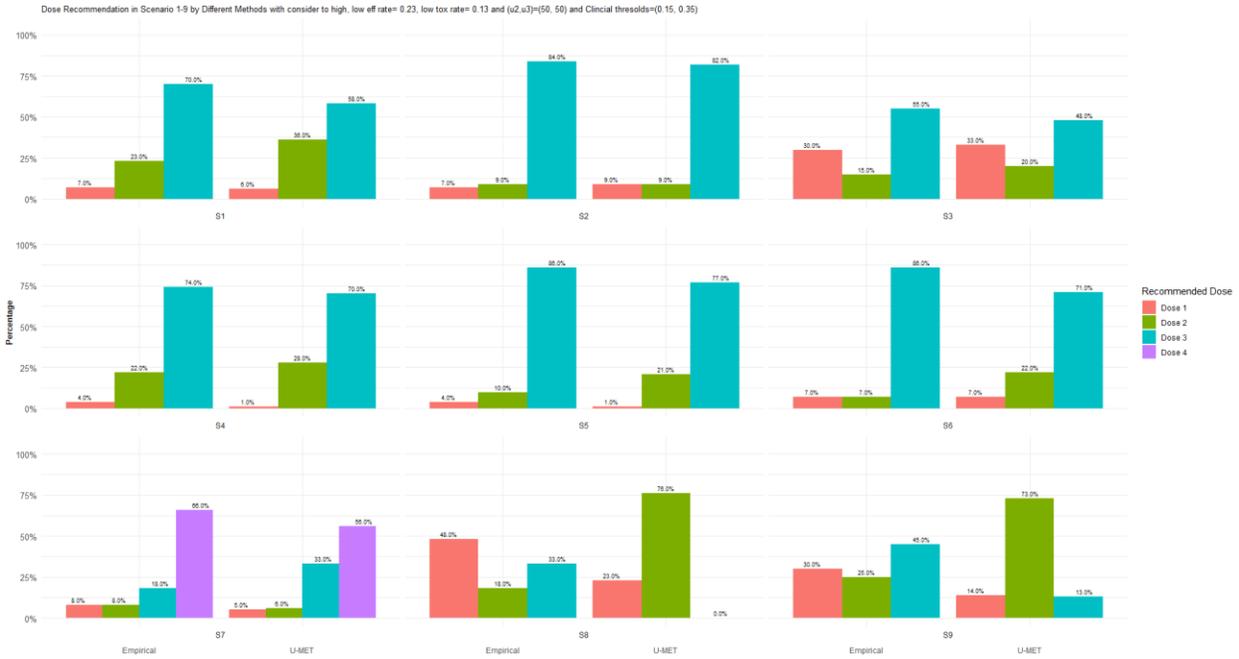

Figure S9 Simulation results of dose selection for n=30, low efficacy rate=0.23, low toxicity rate=0.13, $(u_2, u_3)$=(50, 50), $(ED_1, ED_2)$ = (0.15, 0.35), $\alpha_1 = 0.34$, consider is high dose; (Same set-up as Table 6)

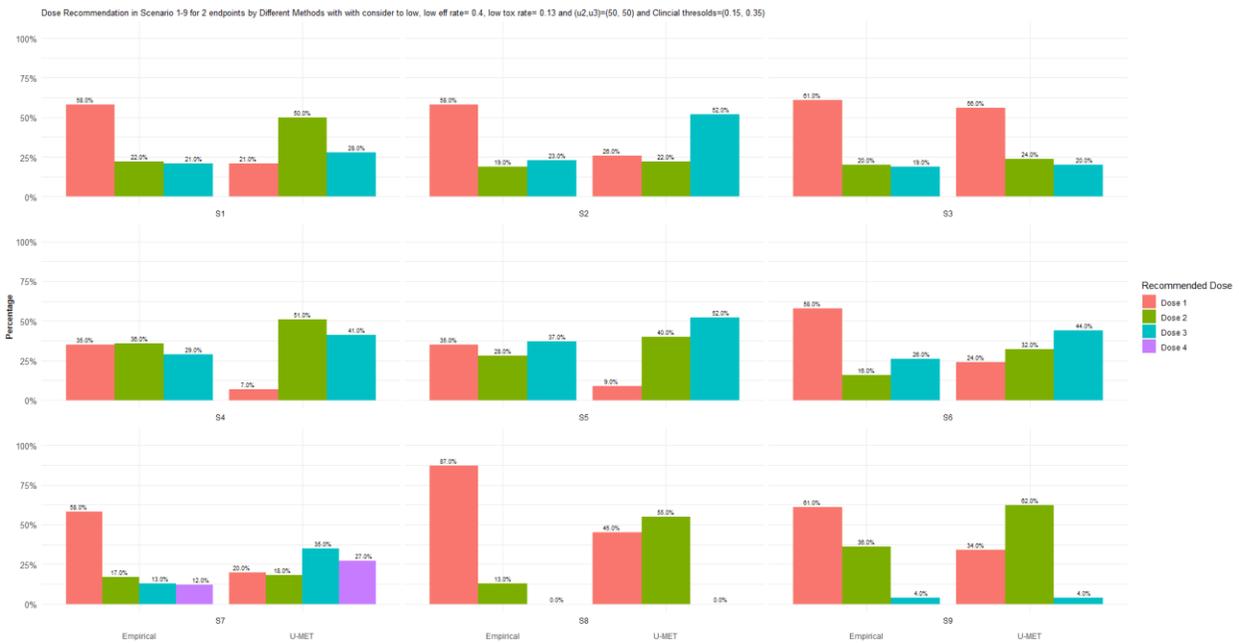

Figure S10 Simulation results of dose selection for n=30, low efficacy rate=0.40, low toxicity rate=0.13, , $(u_2, u_3)$=(50, 50), $(ED_1, ED_2)$ = (0.15, 0.35), $\alpha_1 = 0.20$, consider is low dose (Same set-up as Table S2)

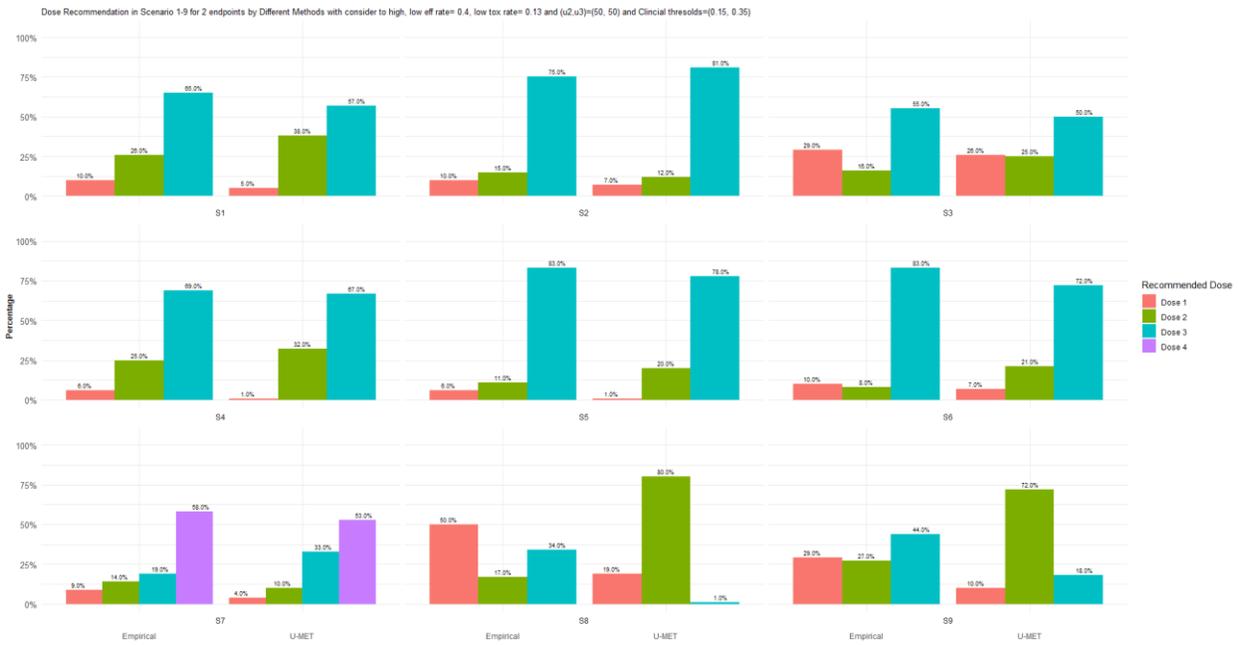

Figure S11 Simulation results of dose selection for n=30, low efficacy rate=0.40, low toxicity rate=0.13, $(u_2, u_3)$=(50, 50), $(ED_1, ED_2)$ = (0.15, 0.35), $\alpha_1 = 0.34$, consider is high dose (Same set-up as Table S2)

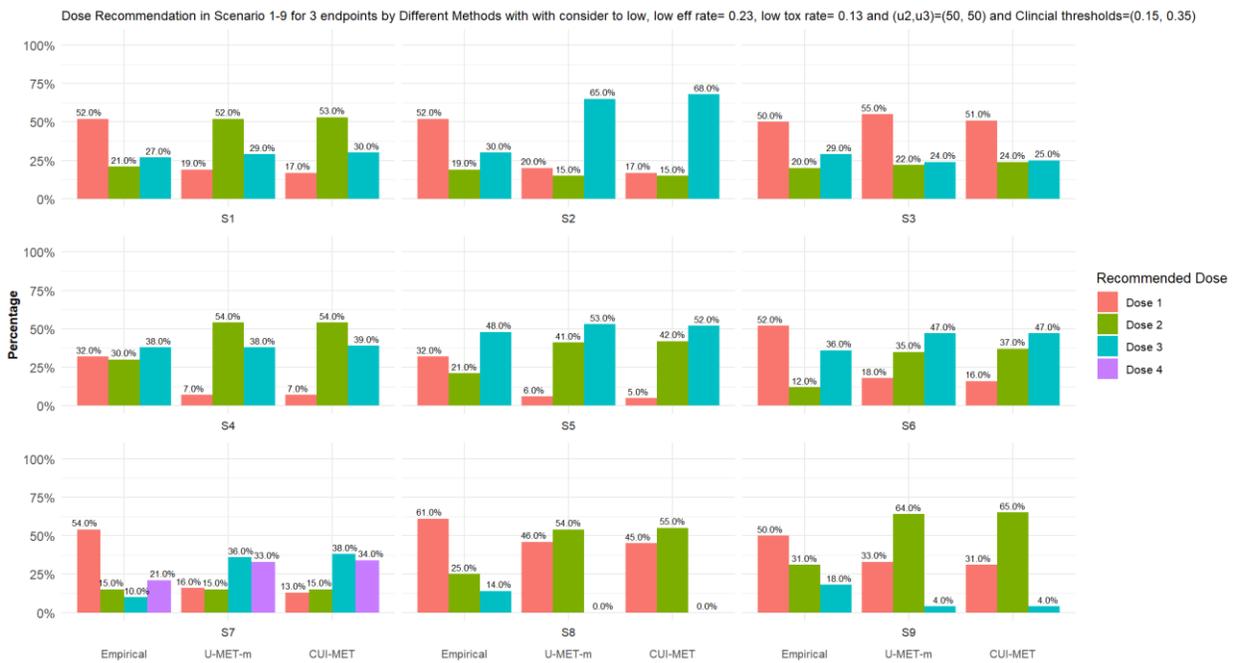

Figure S12 Simulation results of dose selection for n=30, low efficacy rate=0.23, low toxicity rate=0.13, $(u_2, u_3)$=( 50, 50), $(ED_1, ED_2)$ = (0.15, 0.35), $\alpha_1$=0.20, consider is low dose. (same set-up as Table 7)

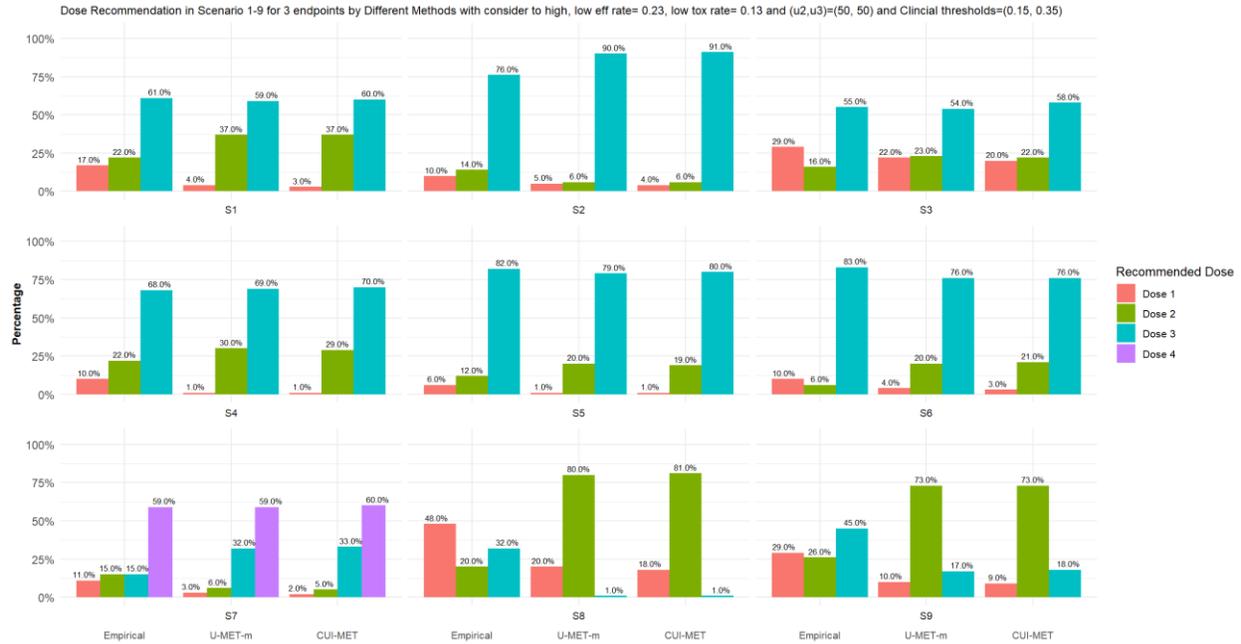

*Figure S13  Simulation results of dose selection for n=30, low efficacy rate=0.23, low toxicity rate=0.13, ($u_2$, $u_3$)=( 50, 50), ($ED_1$, $ED_2$) = (0.15, 0.35), $\alpha_1$=0.34, consider is high dose. (same set-up as Table 7)*

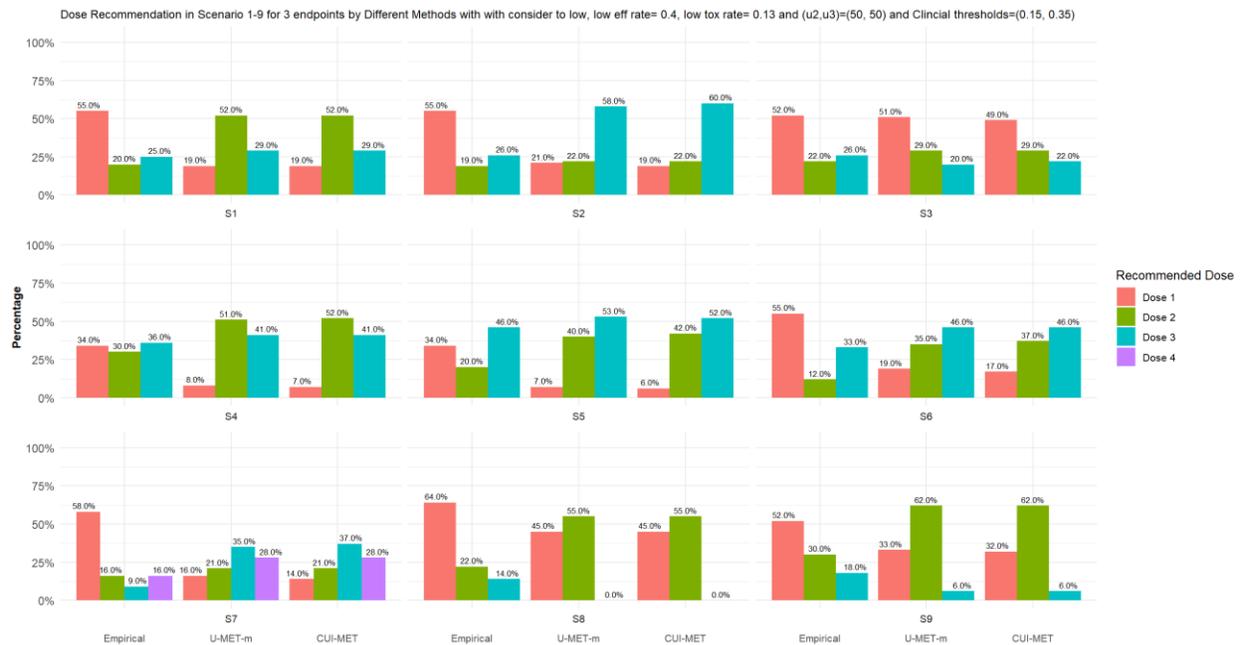

*Figure S14  Simulation results of dose selection for n=30, low efficacy rate=0.40, low toxicity rate=0.13, ($u_2$, $u_3$)=(50, 50), ($ED_1$, $ED_2$) = (0.15, 0.35), $\alpha_1$=0.20, consider is low dose. (same set-up as Table S3)*

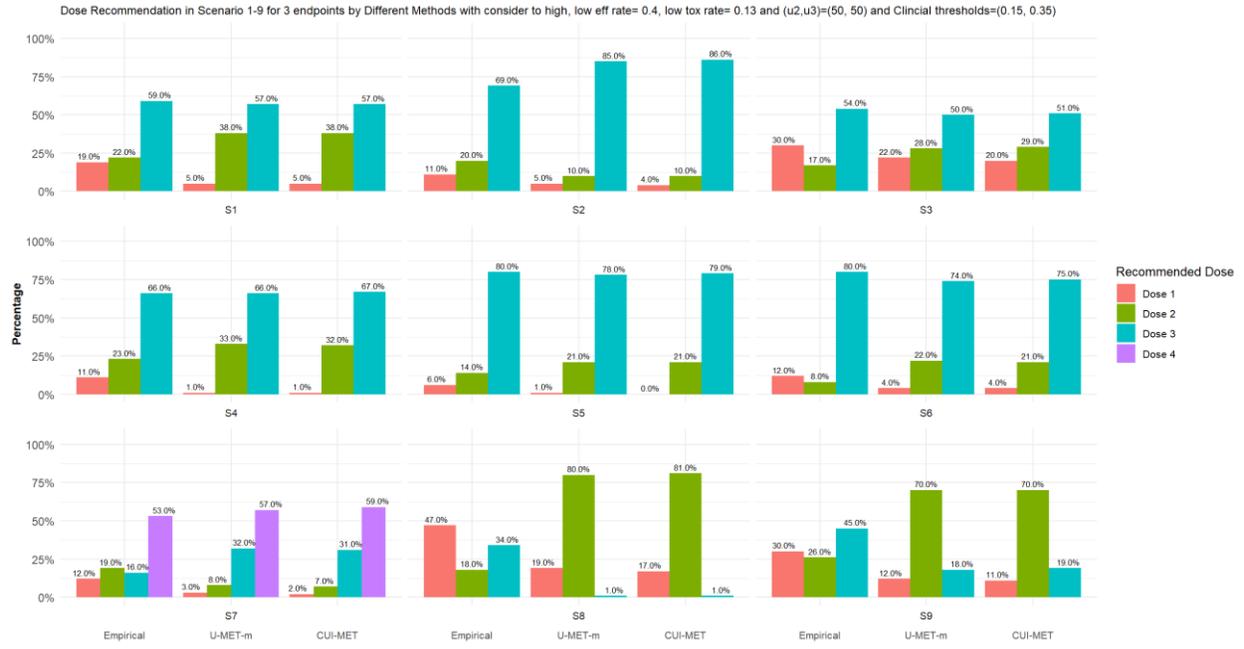

*Figure S15  Simulation results of dose selection for n=30, low efficacy rate=0.40, low toxicity rate=0.13, $(u_2, u_3)$=(50, 50), $(ED_1, ED_2)$ = (0.15, 0.35), $\alpha_1$=0.34, consider is high dose. (same set-up as Table S4)*

Additional analyses for the second simulation setup were run using different utility scores for Table 1, where ($u_2 = 65, u_3 = 35$). The utility score was changed to put more weight on having no toxicity and no efficacy to demonstrate what happens when a shift is made from efficacy and toxicity to no efficacy and no toxicity. Results are shown in Figures S16–S23.

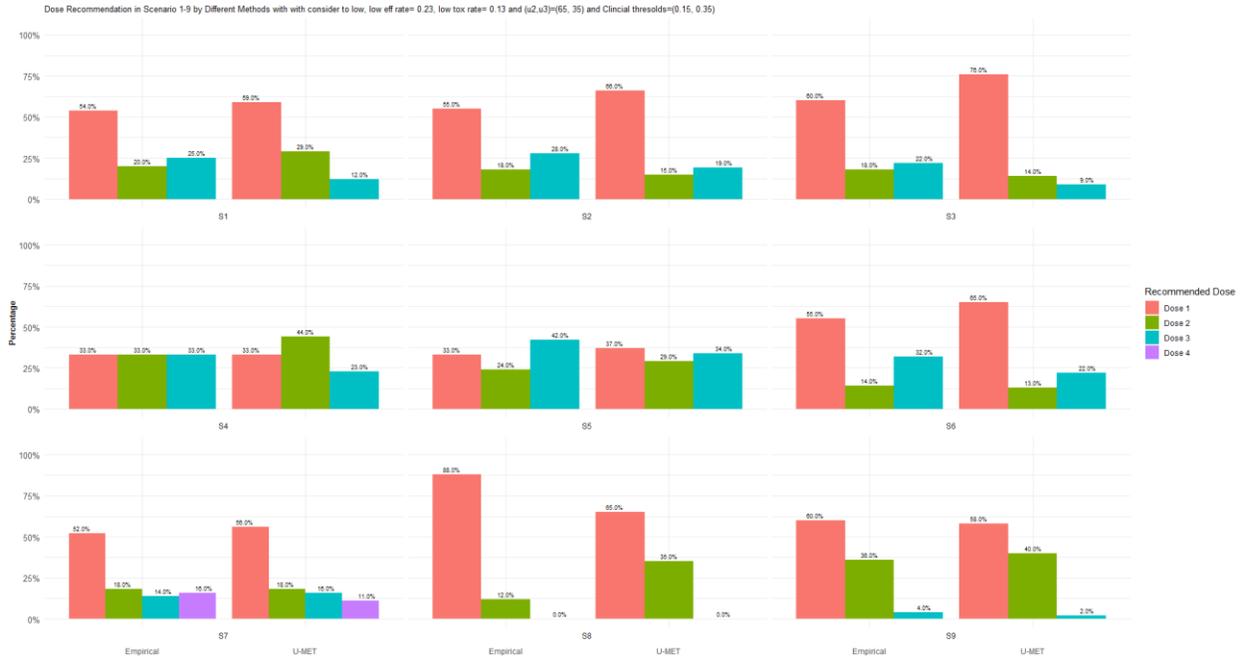

*Figure S16 Simulation results of dose selection for n=30, low efficacy rate=0.23, low toxicity rate=0.13, ($u_2$, $u_3$)=(65, 35), ($ED_1$, $ED_2$) = (0.15, 0.35), $\alpha_1 = 0.20$, consider is low dose (Same set-up as Table 6)*

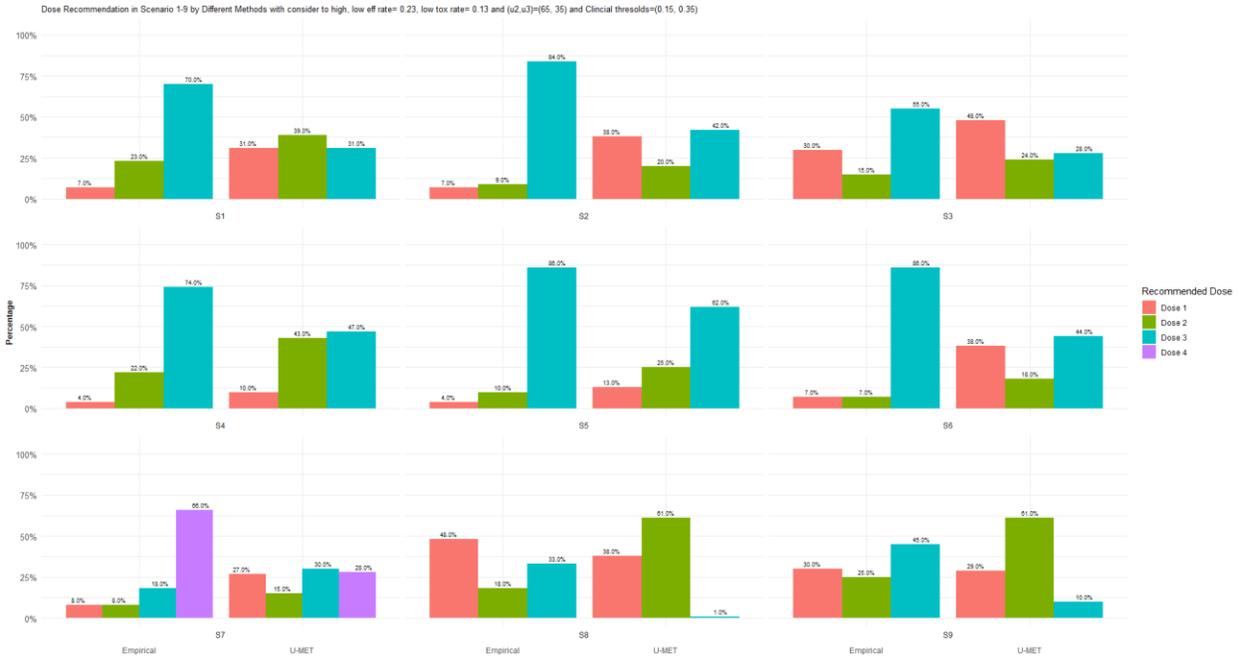

*Figure S17 Simulation results of dose selection for n=30, low efficacy rate=0.23, low toxicity rate=0.13, ($u_2$, $u_3$)=(65, 35), ($ED_1$, $ED_2$) = (0.15, 0.35), $\alpha_1 = 0.34$, consider is high dose (Same set-up as Table 6)*

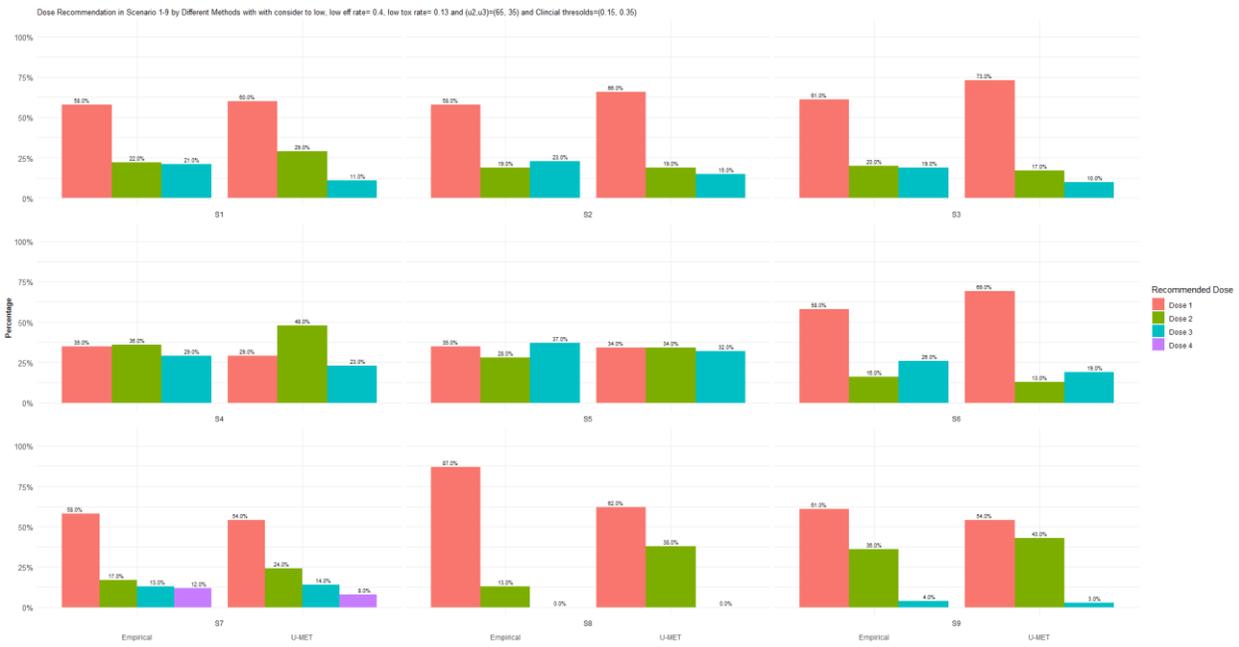

*Figure S18 Simulation results of dose selection for n=30, low efficacy rate=0.40, low toxicity rate=0.13, ($u_2$, $u_3$)=(65, 35), ($ED_1$, $ED_2$) = (0.15, 0.35), $\alpha_1 = 0.20$, consider is low dose (Same set-up as Table S2)*

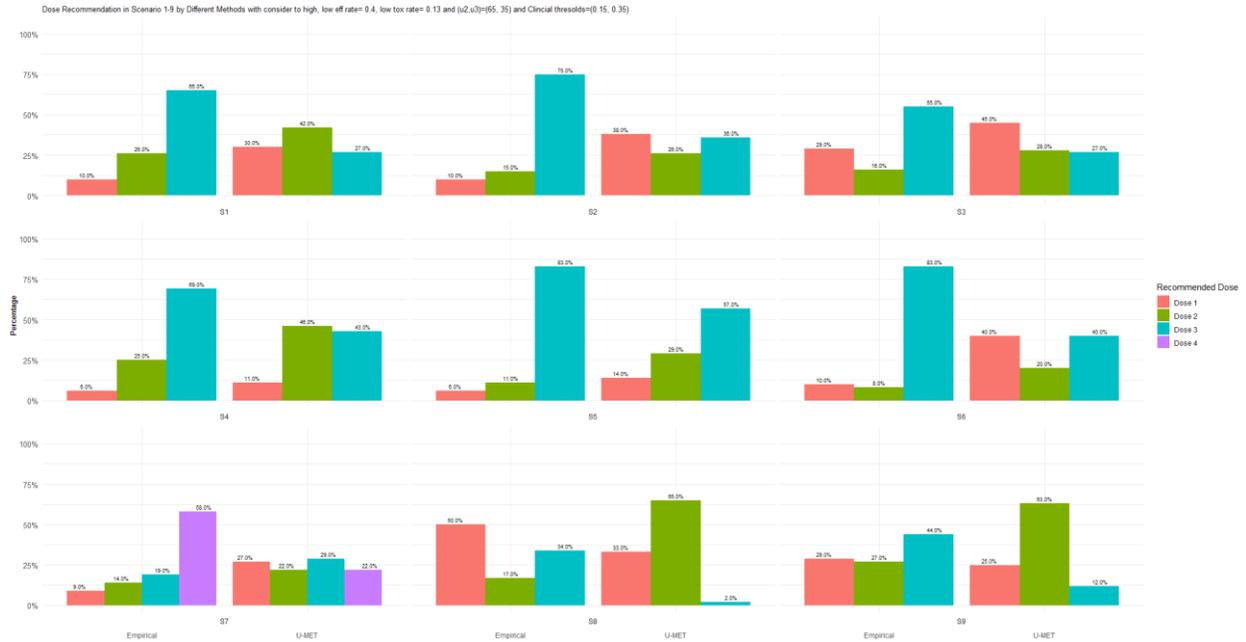

*Figure S19 Simulation results of dose selection for n=30, low efficacy rate=0.40, low toxicity rate=0.13, $(u_2, u_3)$=(65, 35), $(ED_1, ED_2)$ = (0.15, 0.35), $\alpha_1 = 0.34$, consider is high dose (Same set-up as Table S2)*

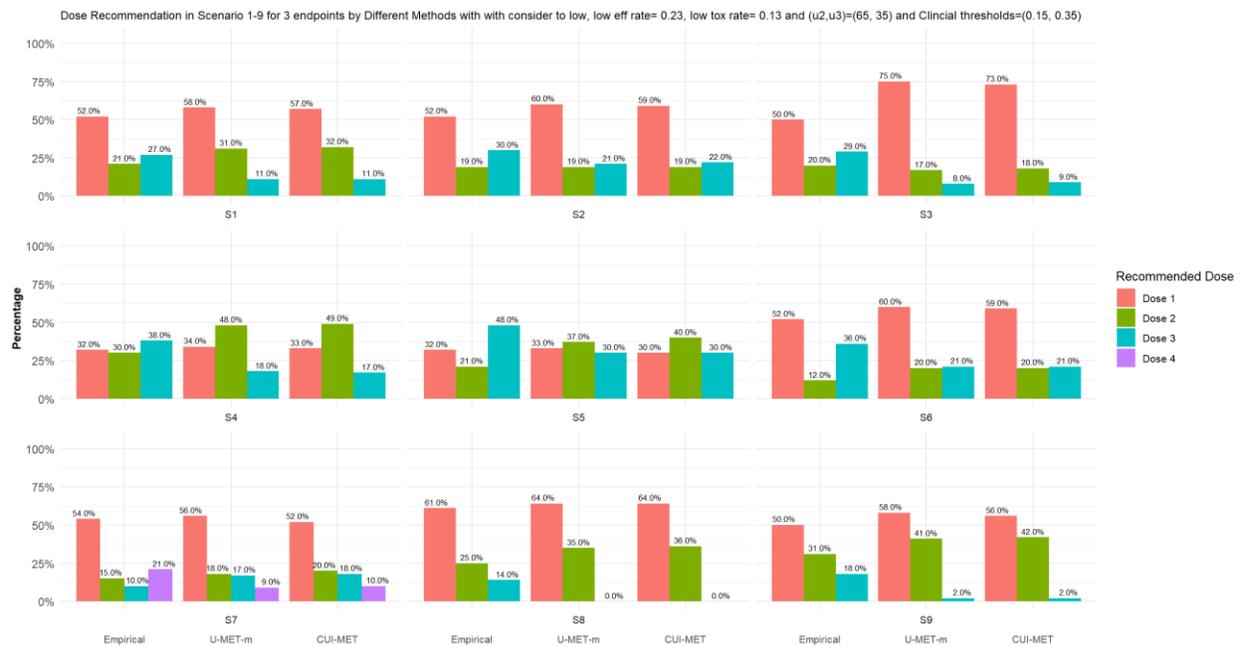

*Figure S20 Simulation results of dose selection for n=30, low efficacy rate=0.23, low toxicity rate=0.13, $(u_2, u_3)$=( 65, 35), $(ED_1, ED_2)$ = (0.15, 0.35), $\alpha_1$=0.20, consider is low dose. (same set-up as Table 7)*

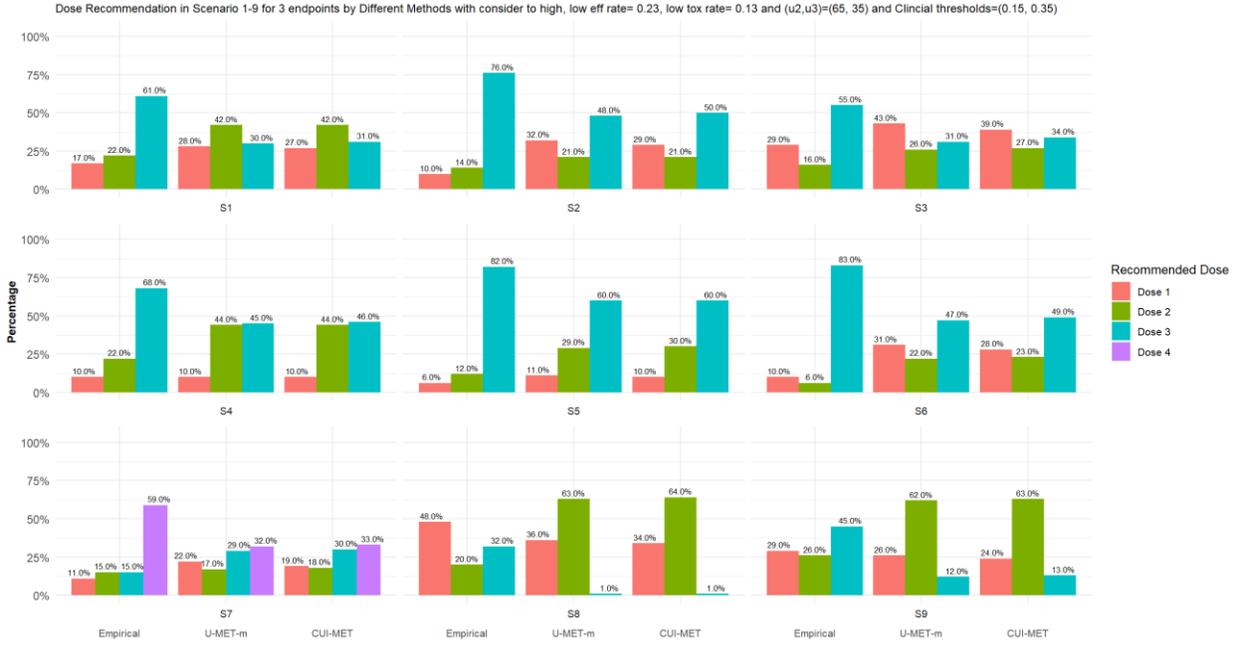

*Figure S21 Simulation results of dose selection for n=30, low efficacy rate=0.23, low toxicity rate=0.13, $(u_2, u_3)=(65, 35)$, $(ED_1, ED_2) = (0.15, 0.35)$, $\alpha_1$=0.34, consider is high dose. (same set-up as Table 7)*

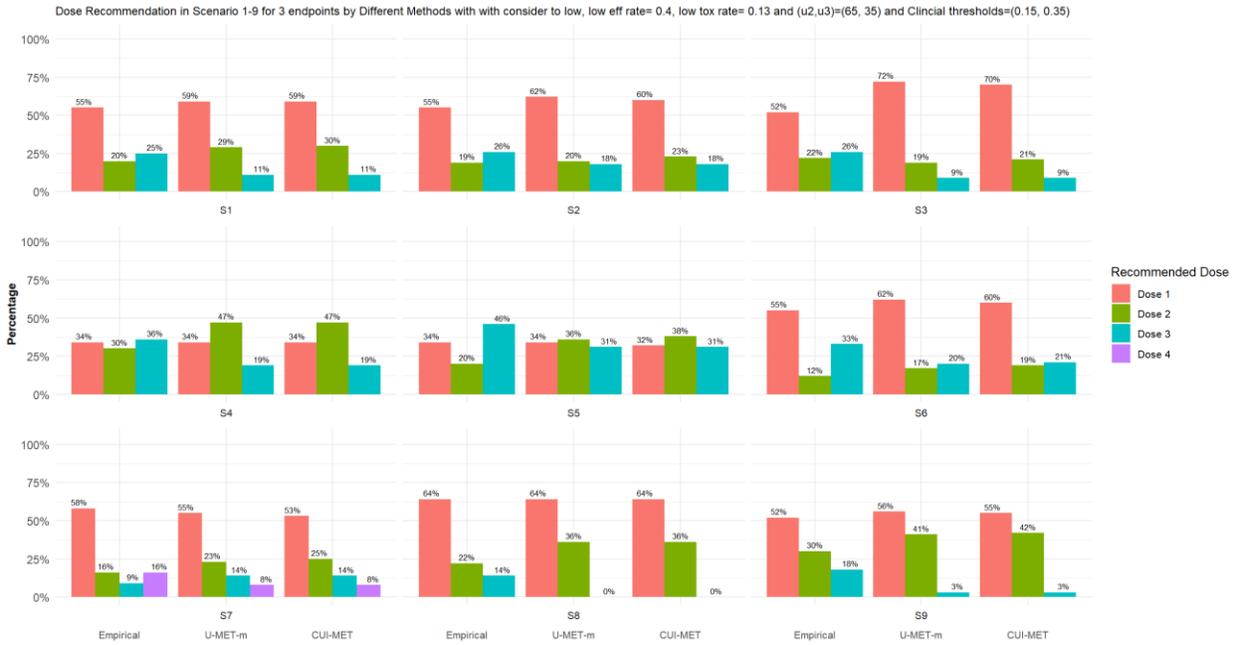

*Figure S22 Simulation results of dose selection for n=30, low efficacy rate=0.40, low toxicity rate=0.13, $(u_2, u_3)=(65, 35)$, $(ED_1, ED_2) = (0.15, 0.35)$, $\alpha_1$=0.20, consider is low dose. (same set-up as Table S3)*

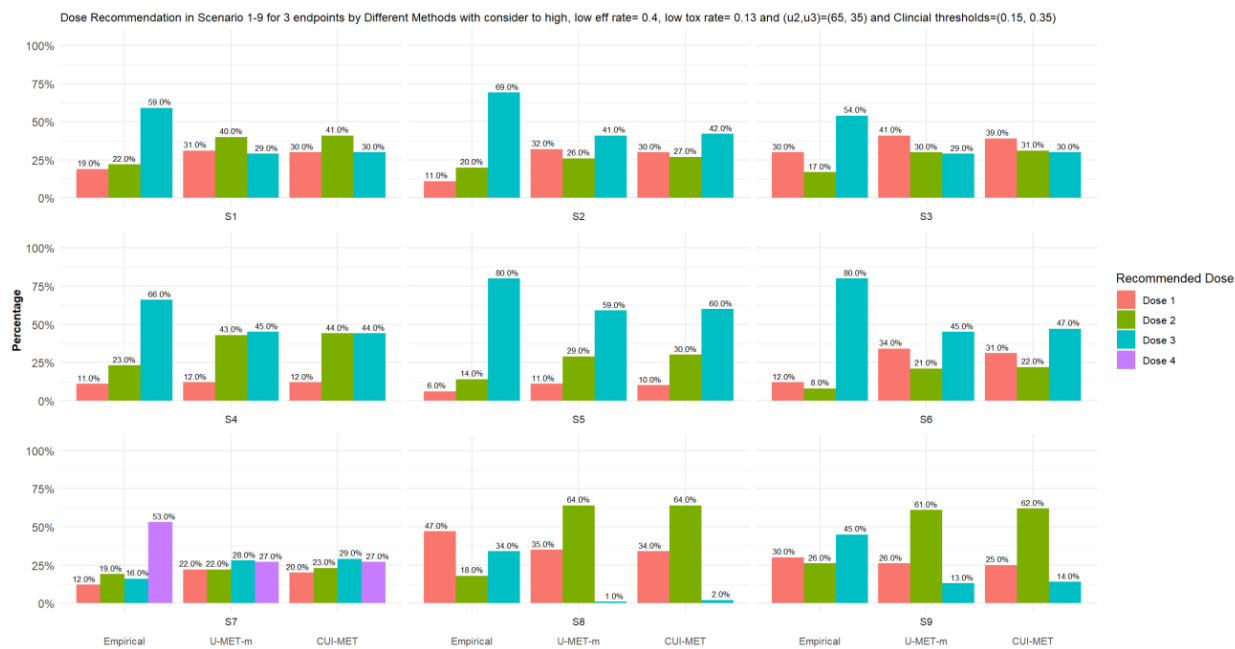

*Figure S23 Simulation results of dose selection for n=30, low efficacy rate=0.40, low toxicity rate=0.13, ($u_2$, $u_3$)=( 65, 35), ($ED_1$, $ED_2$) = (0.15, 0.35), $\alpha_1$=0.34, consider is high dose. (same set-up as Table S4)*

Additional analyses for the examples are provided in Table S5 to report the pairwise analysis.

| Table S5 Results of dose comparison from examples for pairwise test | | | | |
|---|---|---|---|---|
| | Response (%) | Toxicity (%) | Empirical design | U-MET-m |
| Scenario | | | ED, TR, (L,C,H) | $\hat{u}_\Delta^*$, $\Pr(\hat{u}_\Delta^* > 0)$, (L,C,H) |
| 1 (Dose 1, 2, 3) | (47, 57, 76) | (17, 20, 26) | | |
| 1, Dose 1 vs 3 | 47 vs 76 | 17 vs 26 | 29%, 1.53, C | 13.8, 0.870, H |
| 1, Dose 2 vs 3 | 57 vs 76 | 20 vs 26 | 19%, 1.3, H | 9, 0.773, C |
| 1, Dose 1 vs 2 | 47 vs 56 | 17 vs 20 | 10%, 1.18, C | 4.8, 0.648, L |
| 2 (Dose 1, 2, 3) | (47, 67, 60) | (17, 20, 26) | | |
| 2, Dose 1 vs 3 | 47 vs 60 | 17 vs 26 | 13%, 1.53, L | 5.2, 0.657, L |
| 2, Dose 2 vs 3 | 67 vs 60 | 20 vs 26 | -7%, 1.33, L | High dose dropped, L |
| 2, Dose 1 vs 2 | 47 vs 67 | 17 vs 20 | 20%, 1.18, H | 10.80, 0.808, H |
| Abbreviations: C = Consider zone; ED = efficacy rate difference; H = high dose; L = low dose; TR = toxicity ratio. | | | | |